\documentclass[pre,twocolumn,aps,floatfix,10pt]{revtex4}
\usepackage{amsmath}
\usepackage{amsfonts}
\usepackage{amssymb}
\usepackage{graphicx}
\usepackage{fontenc}
\usepackage{psfrag,layout}
\usepackage{upgreek}

\begin{document}
\title{Wall-liquid and wall-crystal interfacial free energies via
thermodynamic integration: A molecular dynamics simulation study}
\author{Ronald Benjamin{$^{1,2}$} and J{\"u}rgen Horbach{$^{2}$}}
\affiliation{$^{1}$Institut f{\"u}r Materialphysik im Weltraum, 
Deutsches Zentrum f{\"u}r Luft- und Raumfahrt (DLR),
51170 K{\"o}ln, Germany\\
$^{2}$Institut f{\"u}r Theoretische Physik II, Universit{\"a}t D{\"u}sseldorf,
Universit\"atsstra\ss e 1, 40225 D{\"u}sseldorf, Germany}

\begin{abstract}
A method is proposed to compute the interfacial free energy of a
Lennard-Jones system in contact with a structured wall by molecular
dynamics simulation. Both the bulk liquid and bulk face-centered-cubic
crystal phase along the (111) orientation are considered. Our approach
is based on a thermodynamic integration scheme where first the bulk
Lennard-Jones system is reversibly transformed to a state where it
interacts with a structureless flat wall. In a second step, the
flat structureless wall is reversibly transformed into an atomistic
wall with crystalline structure.  The dependence of the interfacial free
energy on various parameters such as the wall potential, the density and
orientation of the wall is investigated.  The conditions are indicated
under which a Lennard-Jones crystal partially wets a flat wall.
\end{abstract}

\maketitle

\section{Introduction}
Knowledge of the interfacial free energy between a
crystal or liquid in contact with a solid wall is crucial
to the understanding of heterogeneous nucleation and wetting
phenomena~\cite{gibbs57,adamson97,dietrich88,rowlinson82,degennes85}.
However, interfacial free energies are hardly accessible in experiments
and in fact only a few measurements have been reported so far~(see
e.g.~\cite{adamson97,navascues79,howe97}).

Due to the lack of experimental data, particle-based simulation
techniques such as Molecular Dynamics (MD) and Monte Carlo
(MC)~\cite{allen-tildesley87,landau-binder00} are of special
 importance to understand
the properties of wall-liquid and wall-crystal interfaces
and to rationalize calculations in the framework of density
functional theory~\cite{loewen94evans90,barker76,velasco89}.
In this context, MC and MD simulations have been used to
understand the microscopic mechanism of fluid wetting on solid
surfaces~\cite{navascues79,henderson84,miguel06,nijmeijer90,tang95}
as well as the wetting and drying transition of a fluid
at liquid-vapor coexistence and in contact with a solid
wall~\cite{nijmeijer90,tang95,vanleeuwen90,bruin91}.  The question of how
the wall structure affects the interfacial tension with respect to liquid,
vapor and solid phases has been also addressed~\cite{tang95,leroy09}.

On a macroscopic scale, a crystal that partially wets a wall might
be described as a spherical cap. Then, the contact angle $\theta_{c}$
of the cap with the wall is given by Young's equation~\cite{degennes85},
\begin{equation}
\gamma_{\text{wc}}+\gamma_{\text{cl}}
\cos \theta_{\text{c}}=\gamma_{\text{wl}}\,
\label{eq:young}
\end{equation}
with $\gamma_{\text{wc}}$ the wall-crystal, $\gamma_{\text{cl}}$ the
crystal-liquid, and $\gamma_{\text{wl}}$ the wall-liquid interfacial
free energy. Equation (\ref{eq:young}) describes the condition of a
spherical crystal droplet resting on a wall, being in coexistence with
the liquid phase.  Incomplete wetting corresponds to
contact angles $0<\theta_{c}<\pi$.

On a nanoscopic scale, deviations from Young's equation can
be expected, e.g.~due to the contribution of line tension
effects~\cite{gibbs57,degennes85,weijs11}. To quantify the
latter deviations, reliable estimates of $\gamma_{\text{wc}}$,
$\gamma_{\text{cl}}$ and $\gamma_{\text{wl}}$ are required. Then, the
contact angle can be obtained via Eq.~(\ref{eq:young}) and compared to
a direct measurement of $\theta_{c}$.

In this paper, we propose a thermodynamic
integration (TI)~\cite{frenkel-smit02} scheme for the calculation
of $\gamma_{\text{wc}}$ and $\gamma_{\text{wl}}$.  To obtain
$\gamma_{\text{wl}}$, most previous studies have used the mechanical
approach of calculating the normal and tangential pressure
components at the wall and integrating over the pressure anisotropy
(PA)~\cite{navascues79,henderson84,nijmeijer90,tang95,vanleeuwen90}.
While the PA method is  valid for planar wall-liquid or
liquid-vapor interfaces it fails in case of small liquid drops
in contact with a solid wall~\cite{sampayao10}.  Moreover, its
use is justified only for systems where the interfacial tension
equals the interfacial free energy~\cite{tiller91}. This is
true for a wall represented by a time-independent external
field~\cite{henderson84,varnik00,varnik-thesis00} or a
wall made of particles rigidly fixed at the sites of an ideal
lattice~\cite{nijmeijer90,tang95,vanleeuwen90,bruin91,bakker89,crevecoeur95}.
However, for systems which can support stress, such as a wall
consisting of a {\textquotedblleft fully interacting solid
phase\textquotedblright}~\cite{laird10}, this method is invalid. For
the same reason, the PA technique cannot be used to determine
$\gamma_{\text{wc}}$~\cite{tiller91}.  Even for wall-liquid interfaces,
the PA method can yield results with acceptable precision only with huge
computational effort.  Most previous works based on the PA technique
yielded results of low accuracy and the values of the interfacial tension
reported in the literature differ widely, even for simple systems.

Due to the obvious disadvantage in using the PA method, a few
thermodynamic approaches have  been developed to evaluate the wall-liquid
and wall-crystal interfacial free energies with improved precision.
Heni and L{\"o}wen~\cite{heni99} combined MC simulations and thermodynamic
integration to determine the interfacial free energies of hard sphere
liquids and solids near a planar structureless wall over a whole range
of bulk densities including the solid-liquid coexistence density.
In their thermodynamic integration scheme, a bulk hard sphere system
was reversibly transformed into a system interacting with a more
and more impenetrable wall and finally a hard wall. Fortini
and Djikstra~\cite{fort-djik06} used a thermodynamic integration scheme
based on exponential potentials to calculate $\gamma_{\text{wl}}$ and
$\gamma_{\text{wc}}$  at bulk coexistence conditions. Their results
were in good agreement with those of Heni and L{\" o}wen but obtained
with significantly higher precision. Due to precrystallization of the
hard spheres near the wall close to the bulk freezing transition, both
Heni and L{\"o}wen and Fortini and Dijkstra extrapolated the value of
the interfacial tensions at coexistence from the data at lower densities.

Laird and Davidchack~\cite{laird07} developed a TI
method by the use of ``cleaving potentials", to obtain
$\gamma_{\text{wl}}$ and $\gamma_{\text{wc}}$ for hard sphere systems at
coexistence. In another work~\cite{laird10}, they used the ``Gibbs-Cahn
integration" method, to obtain wall-fluid interfacial free energies
for hard sphere systems. This method yielded results consistent with
the TI method with ``cleaving potentials" but
were obtained with significantly less computational effort. However,
``Gibbs-Cahn integration" requires that one knows already the interfacial
free energy at one point.  Deb {\it et al.}~\cite{deb10,deb11} compared
different methods to obtain wall-fluid and wall-crystal interfacial free
energies for hard sphere systems confined by hard walls or, soft walls
described by the Weeks-Chandler-Anderson (WCA) potential. They introduced
a scheme similar to Wang-Landau sampling \cite{wang01}, known as the
``ensemble mixing" method, to perform a TI from
a system without walls to a system confined by walls. For hard spheres,
Deb {\it et al.} obtained good agreement with the results of Laird
and Davidchack.

In contrast to these few works on hard-sphere systems, there is a
dearth of results on the interfacial free energies of systems with
continuous potentials, such as Lennard-Jones (LJ) systems.  Recently, Leroy {\it et
al.}~\cite{leroy09} obtained $\gamma_{\text{wl}}$  for a LJ liquid in
contact with a flexible LJ structured wall by the use of a TI
technique, known as the ``phantom wall'' method. In this
approach, the structured wall interacting with the liquid is gradually
moved away from the liquid, while a  structureless flat wall is moved
towards it such that in the final state, the liquid interacts only with
the structureless wall. Computing the free energy difference during this
transformation, along with the interfacial free energy of liquid in contact
with the structureless wall, gives $\gamma_{\text{wl}}$. $\gamma_{\text{wl}}$ 
for the liquid-flat wall system, which serves as the reference state for 
their system was obtained using the PA technique.  Since the PA 
technique fails in case of crystal-wall interfaces~\cite{tiller91} one cannot 
use their scheme to determine  $\gamma_{\text{wc}}$ for crystal in contact 
with a structured wall. In fact, much less is known  about 
$\gamma_{\text{wc}}$ for LJ systems in contact with a wall.

Grochola {\textit et al.}~\cite{grochola02ab0405ab} developed another
TI technique which they have called ``$\lambda$-integration", to determine
the surface free energies of solids. In principle,
this technique could be also applied to wall-crystal or wall-liquid interfaces,
but the method has not been worked out yet for such interfaces.

A straightforward and comprehensive method is thus needed to compute the
interfacial free energies of LJ systems in contact with a wall.  In the
present work, a novel TI scheme is introduced
to compute the interfacial free energy of a LJ system confined between
walls. We consider both the liquid as well as the fcc crystal phase
along the (111) orientation near the wall.  While most previous works
employing TI methods to obtain $\gamma_{\text{wl}}$
or $\gamma_{\text{wc}}$ are limited to structureless walls, here
we specifically consider the case of a structured wall, consisting
of particles rigidly attached to the sites of an ideal fcc lattice.
Our scheme consists of TI in two steps, providing
a reversible thermodynamic path that transforms the bulk LJ system into
a LJ fluid or crystal interacting with a structured wall.  In the first
step, a thermodynamic path is devised to reversibly transform the bulk LJ
system without walls and periodic boundary conditions in all directions
to a state where it interacts with the structureless wall. This is
accomplished by gradually modifying the interaction potential between the
wall and the LJ particles along the thermodynamic path.  The technique
is inspired by the method proposed by Heni and L{\"o}wen~\cite{heni99}
to compute the interfacial free energy of hard sphere fluids and crystals
in contact with a hard wall.

The LJ system interacting with the flat wall serves as the
reference state to calculate the interfacial free energy of the LJ liquid
or crystal in contact with a structured wall.  In the second step, another
TI scheme reversibly changes the structureless
wall interacting with the LJ system into a structured wall.  This is done
by gradually switching off the flat walls and simultaneously switching
on the structured walls.  While previous methods based on TI
techniques for the calculation of $\gamma_{\rm wl}$ make use
of {\textquotedblleft cleaving potentials\textquotedblright}~\cite{laird07}
or {\textquotedblleft phantom walls \textquotedblright}~\cite{leroy09}, 
here we directly modify the interaction potential to make the transformation
from the reference state to the final state in each of the two steps.
Though this TI scheme is specifically developed for
a LJ potential, it can be easily generalized to more complex potentials.

The wetting  behavior of a liquid or crystal in contact with a structured
solid wall will be affected by various parameters. In this work we focus
on three parameters: i) the interaction strength between the wall and
the LJ system, ii) the density of the structured wall, and iii) the
orientation of the structured wall with respect to the interface normal.
For these cases, wall-crystal interfacial free energies only for the (111)
orientation of the crystal are considered.  

Since the PA method has been
widely applied in the past to evaluate the wall-liquid interfacial free
energy, we will compare results obtained from it with those yielded by
the TI method for both flat and
structured walls. In addition, we will also show that the interfacial free
energies of the LJ system interacting with a flat wall can be obtained
directly at coexistence, without any extrapolation from data at low
densities, enabling us to investigate its wetting behavior.

Furthermore, to compare the estimates of interfacial free energy
yielded by our TI scheme with that obtained
in previous works, we apply our technique to a model system  studied
by Tang and Harris (TH)~\cite{tang95} using the mechanical definition
of the interfacial free energy.  Their system consisted of a LJ fluid
confined between identical rigid structured walls oriented along the
(100) orientation, under conditions of liquid-vapor coexistence.  Later,
Grzelak and Errington (GE)~\cite{grzelak08} investigated the same system using
Grand Canonical Transition Matrix Monte Carlo (GCTMMC) simulations. They
computed the interfacial free energy profile as a function of the
surface density at bulk liquid-vapor saturation condition, to obtain the
contact angle and the solid-vapor and solid-liquid interfacial tensions.
For this system, we will examine the variation of $\gamma_{\rm wl}$ as
a function of the wall-liquid interaction strength and compare estimates
of $\gamma_{\rm wl}$ from these two studies.  Due to the paucity of
studies on the crystal-wall interfacial free energy, we will restrict
this comparison with previous works only to the wall-liquid interfacial
free energy.

In the following, we introduce the details of the model potentials
considered in this work (Sec.~\ref{sec_mp}), give the various
definitions of interfacial free energies, outline the PA method,
describe the proposed TI scheme, and provide the main details
of the simulation (Sec.~\ref{sec_calc}).  Then, we present the
results (Sec.~\ref{sect:results}) and finally draw some conclusions
(Sec.~\ref{sec_conc}).

\section{Model Potential}
\label{sec_mp}
The MD system to determine the interfacial excess free energy of a LJ
system in contact with a structured wall consists of $N$ identical
particles interacting with each other and with the structured wall via
a shifted-force LJ~\cite{allen-tildesley87} potential. If two particles $i$ and
$j$ of types $\alpha$ and $\beta$ are separated by a distance $r_{ij}$,
the interaction potential is written as
\begin{equation}
u_{\alpha\beta}({r_{\text{ij}}}) =
\begin{cases}
\phi_{\alpha\beta}(r_{ij})-\phi_{\alpha\beta}(r^{\text{c}})- 
\phi_{\alpha\beta}^{\prime}(r_{ij}=r^{c})[r_{ij}-r^{\text{c}}] \\
\qquad \qquad \text{for $0 < r_{ij} \leq  r^{\text{c}}$},
\\
 0  \qquad \qquad  \text{for $r_{ij}> r^{\text{c}} $},
\end{cases}
\label{eq:sflj}
\end{equation}
where the prime in $\phi_{\alpha\beta}^{\prime}$ denotes the derivative
with respect to $r$ and
\begin{equation}
\label{eq:lj}
\phi_{\alpha\beta}(r_{ij})=
4\epsilon_{\alpha\beta}
\left[\left(\frac{\sigma_{\alpha\beta}}{r_{ij}}\right)^{12}
-\left(\frac{\sigma_{\alpha\beta}}{r_{ij}}\right)^{6}
\right].
\end{equation}
In Eq.~(\ref{eq:sflj}), $\alpha$ or $\beta$ can represent a LJ
particle (p) or a structured wall particle (w).  The parameters
$\epsilon_{\alpha\beta}$ and $\sigma_{\alpha\beta}$ have units of
energy and length, respectively. The cut-off distance is set to
$r^{c}=2.5\sigma_{\alpha\beta}$.

In the following, energies, lengths and masses are given
in units of $\epsilon_{\text{pp}}$,  $\sigma_{\text{pp}}$
and $m_{\text{p}}$, respectively. Thus, temperature,
pressure and interfacial free energy are expressed in units of
$\epsilon_{\text{pp}}/k_{B}$, $\epsilon_{\text{pp}}/\sigma_{\text{pp}}^3$
and $\epsilon_{\text{pp}}/\sigma_{\text{pp}}^2$, respectively.  Time is
made dimensionless by reducing it with respect to the characteristic time
scale $\sqrt{m_{\text{p}}\sigma_{\text{pp}}^{2}/\epsilon_{\text{pp}}}$.
For simplicity, we choose $\sigma_{\rm wp}=\sigma_{\rm pp}$.

The $N$ identical liquid or crystal particles are enclosed within
a simulation box of size $L_{\text{x}}\times L_{\text{y}} \times
L_{\text{z}}$, using periodic boundary conditions in the $x$ and
$y$ directions. In the $z$ direction the particles are confined
by the structured wall, between $z=z_{\text{b}}$ at the top and
$z=z_{\text{t}}$ at the bottom.  The system thus consists of two
planar wall-liquid (or wall-crystal) interfaces with a total area
of $A=2L_{\text{x}}L_{\text{y}}$.  The structured wall is arranged
in a manner such that the wall layers closest to the LJ system are
positioned at $z_{\text{b}}=-L_{\text{z}}/2$ and
$z_{\text{t}}=L_{\text{z}}/2$. Also, an integer number of unit cells was
chosen for the structured wall such that the wall is exactly 
adapted to the lateral size of the simulation cell.

The width of the structured wall is chosen large enough to avoid LJ
particles  on opposite sides of the wall from interacting with each other
since the determination of interfacial free energy by TI
or PA methods is built on the assumption
of two independent wall-liquid (or wall-crystal) interfaces.

The TI scheme adopted in this work consists of
two steps. First, a bulk LJ system with periodic boundary conditions is
transformed into a state where the LJ system interacts with impenetrable
flat walls. Then, in the second step, the flat walls are reversibly
transformed into structured walls.  The structureless flat wall (fw)
is taken to be a purely repulsive potential interacting along the $z$
direction with the LJ particles and is described by a WCA potential,
\begin{equation}
 \label{eq:uwallwca}
u_{\text{fw}}(z)=
\begin{cases}
 4\epsilon_{w}\left[\left(\frac{\sigma_{\text{pp}}}{z}
\right)^{12}-\left(\frac{\sigma_{\text{pp}}}{z} \right)^{6}
 +\frac{1}{4}\right] \times w(z) \\
 \qquad \text{for  $0 < z \leq z_{\text{cw}}$},  \\
0 \qquad \text{for $z> z_{\text{cw}}$}
\end{cases}
\end{equation}
with the cut-off $z_{\text{cw}}=2^{1/6}\sigma_{pp}$ and $z=z_{i}-Z$ the
distance of particle $i$ at $z_{i}$ to one of the flat walls at $Z=z_{\rm
b}$ or $Z=z_{\rm t}$.  The function $w(z)$ ensures that $u_{\text{fw}}(z)$
goes smoothly to zero at $z=z_{\text{cw}}$ and is given by
\begin{equation}
 w(z)=\frac{1}{h^{4}+\left(1/(z-z_{\text{cw}})^4\right)},
\end{equation}
where the dimensionless parameter $h$ is set to $0.005$.

To compare to the results of Tang and Harris~\cite{tang95} and Grzelak
and Errington~\cite{grzelak08}, we also consider a truncated and shifted
LJ potential for the particle-particle (pp) and particle-structured wall (pw)
interactions,
\begin{equation}
\label{eq:splj}
u_{\alpha\beta}^{\ast}(r_{\text{ij}})=
\begin{cases}
 \phi_{\alpha\beta}(r_{ij})-\phi_{\alpha\beta}(r^{c})
 \qquad \text{for $ r_{ij} < r^{c}$},  \\
0 \qquad \qquad \qquad \qquad \quad \hspace{0.225cm} \text{for $ r_{ij} \geq r^{c}$}
\end{cases}
\end{equation}
with $\alpha\beta = {\rm pp, pw}$ and the cut-off
radius $r^{c}=2.5\sigma_{\text{pp}}$.  Moreover, we choose
$\sigma_{\text{pw}}=1.1 \sigma_{\text{pp}}$ and vary the parameter
$\epsilon_{\rm pw}$ in units of $\epsilon_{\rm pp}$ in order to determine
the interfacial free energy $\gamma_{\rm wl}$ as a function of the
strength of the pw interactions.  As Tang and Harris~\cite{tang95}, we
use a substrate consisting of three layers of atoms rigidly fixed to fcc
lattice sites, with the (100) orientation of the wall facing the liquid
along the $z$ direction. The average number density of the liquid is set
to $\rho_{\text{p}}=0.661\;\sigma_{\rm pp}^{-3}$ and that of the substrate
to $\rho_{\text{w}}=0.59\;\sigma_{\rm pp}^{-3}$ keeping the temperature
of the system fixed at $T=0.9k_{\text{B}}/\epsilon_{\text{pp}}$.

\section{Calculation of interfacial free energies}
\label{sec_calc}
\subsection{Definitions}

The  Hamiltonian of our model, corresponding to the LJ system interacting
with a solid wall, can be written as
\begin{equation}
\begin{split}
H({\bf r},{\bf p})=  
\sum_{i=1}^{N_{\text{p}}} \frac{1}{2m_{i}}{{\bf p}_{i}^{2}}+
\sum_{i=1}^{N_{\text{p}}}\sum_{j=i+1}^{N_{p}} u_{\text{pp}}({r_{ij}}) +
U_{\text{wall}}
\end{split}
\label{eq:hamilt}
\end{equation}
with $N_{\text{p}}$ the total number of LJ particles and $U_{\text{wall}}$
the wall-particle potential. For interactions of the LJ system with a flat
wall, $U_{\text{wall}}=\sum_{i=1}^{N_{\text{p}}}u_{\text{fw}}(z=z_{i}-Z)$,
and for the system in contact with a structured wall,
$U_{\text{wall}}=\sum_{i=1}^{N_{\text{p}}}\sum_{j=1}^{N_{\text{w}}}u_{\text{pw}}(r_{ij})$
(with $N_{\text{w}}$ the total number of wall particles).  In
Eq.~(\ref{eq:hamilt}), there is no kinetic energy term for the walls,
since the flat walls are considered to be of infinite mass and immovable;
similarly, the structured wall particles are considered to be immobile.

Our simulations are performed in the $NP_{\text{N}}AT$ ensemble, where
the number of particles $N$, surface area $A$ and temperature $T$
are kept constant and the length of the simulation box along the $z$
direction is allowed to fluctuate in order to maintain a constant
normal pressure $P_{\text{N}}$.  The use of the $NP_{N}AT$ ensemble
is necessary to maintain a constant bulk density of the system when
TI is applied (see below).  Moreover, any stress
present in the crystal due to interaction with the walls can relax during
the $NP_{N}AT$ simulation.  The determination of the interfacial free
energy by thermodynamic or mechanical approaches demands that there is
a bulk region in the middle of the simulation box where the density is
equal to the bulk density of the homogeneous system. Hence, the system
size along the $z$ direction must be large enough to prevent the two
walls on either side of the LJ system from influencing each other.

The  isothermal-isobaric partition function corresponding to the
Hamiltonian (\ref{eq:hamilt}) is
\begin{equation}
\begin{split}
Q_{NP_{\text{N}}AT}= &\frac{1}{h^{3N}N!}\int \int \int \exp \left
[-\frac{H({\bf r},{\bf p})
+P_{\text{N}}AL_{\text{z}}}{k_{\text{B}}T}\right]\\
 & \times A dL_{\text{z}} d\textbf{r}^{N} d\textbf{p}^{N}
\label{eq:partfunc}
\end {split}
\end{equation}
where ${\bf r}$ and ${\bf p}$ denote respectively the positions and
momenta of the particles and $h$ is the Planck constant. The Gibbs free
energy $G$ of the confined liquid or crystal is related to the partition
function (\ref{eq:partfunc}) by $G=-k_{\text{B}}T\ln Q_{NP_{\text{N}}AT}$.

The derivative of Gibbs free energy with respect to the surface area
defines the interfacial tension:
\begin{equation}
\label{eq:Gibbs_gamma}
 \gamma^{\prime} = \left ( \frac{\partial G}{\partial A} \right )_{NP_{\text{N}}AT}.
\end{equation}
This thermodynamic definition of the interfacial tension is equivalent
to the mechanical definition~\cite{kirkwood49}:
\begin{equation}
\gamma^{\prime} =\frac{1}{2}
\int_{z_{\text{b}}}^{z_{\text{t}}} [P_{N}(z)-P_{T}(z)]
\label{eq:surften}
 \end{equation}
where $P_{N}(z)$ and $P_{T}(z)$ are respectively the normal and
tangential pressure profiles of the liquid and the factor $1/2$ is
introduced to account for the fact that the liquid is confined between
two identical walls. The local pressure tensor components $P_{N}(z)$
and $P_{T}(z)$ are defined in Eqs.~(\ref{Eq:PNliq}), (\ref{eq:Pwall})
and (\ref{Eq:PTliq}) (see next section).

The interfacial tension $\gamma^{\prime}$ is related to the interfacial
free energy $\gamma$ as~\cite{shuttleworth50}
\begin{equation}
 \gamma^{\prime}=\gamma + A\, \frac{\partial \gamma}{\partial A} .
\label{eq:interftensiondefn}
\end{equation}
If a liquid is in contact with a dynamic structured wall, which can
support stress, the interfacial excess free energy will vary with
the area of the interface. However, in this work we consider rigid
substrates and structureless flat walls, which do not support stress
and hence for the liquid-wall interface, the interfacial tension will
be equal to the interfacial free energy validating the use of the
PA method.  For a crystal-wall interface, however, the second term in
Eq.~(\ref{eq:interftensiondefn}) will be a relevant quantity.  In this
work, we will restrict our attention only to the determination of the
interfacial free energy.

The interfacial free energy of an inhomogeneous system with walls can be
defined as a Gibbs excess free energy per area,
\begin{equation}
\label{eq:defife}
 \gamma = \frac{G_{\text{system}} - G_{\text{bulk}}}{A}
\end{equation}
with $G_{\text{system}}$ and $G_{\text{bulk}}$ the Gibbs free energies
of the inhomogeneous system and the bulk phase of the system,
respectively. We will use this  definition to calculate the interfacial
free energy using TI.

\subsection{$\gamma$ from PA}
\label{sec:PA}
Determination of the interfacial free energy by the PA method is
only valid if the interfacial tension equals the interfacial free
energy. This holds, e.g., for interfaces between a liquid and a flat
wall or rigid substrate.  Hence, we will use the PA technique to obtain
the wall-liquid interfacial free energy, and compare it with results
obtained from TI.

To obtain interfacial free energies from the mechanical approach,
the local tangential and normal pressure tensor components have to
be computed.  There is no unique microscopic definition for these
local pressure tensor components and different expressions lead to
the same value for the interfacial tension~\cite{varnik00}.  Mechanical
stability, however, requires that the normal component of the pressure
tensor is independent of the distance from the wall and furthermore
the two tangential components along the $x$ and $y$ directions are
equal to each other. In the literature, it is only the Irving and
Kirkwood (IK) definition of the pressure tensor that satisfies these
properties~\cite{irving50,varnik00,varnik-thesis00}.  According to the IK definition,
contributions to the normal and tangential components of the pressure
tensor from any two particles $i$ and $j$ at $z_{i}$ and $z_{j}$,
respectively, can be written as
\begin{align}
 P_{\text{N}}^{\text{IK}}(z) & =\rho(z)k_{\text{B}}T  \nonumber  \\
& -\frac{1}{A} \left\langle
\sum_{i<j}\frac{z_{ij}}{r_{ij}} u^{\prime}_{\text{pp}}(r_{ij})
\Theta
\left(\frac{z-z_{i}}{z_{ij}}\right)
\Theta\left(\frac{z_{j}-z}{z_{ij}}\right)\right\rangle \nonumber  \\
&   -\frac{1}{A} \left\langle
\sum_{i=1}^{N_{\text{p}}}\sum_{j=1}^{N_{\text{w}}}
\frac{z_{ij}}{r_{ij}} u^{\prime}_{\text{pw}}(r_{ij})
\Theta \left(\frac{z-z_{i}}{z_{ij}}\right)
\Theta\left(\frac{z_{j}-z}{z_{ij}}\right)\right\rangle 
\label{Eq:PNliq}
\end{align}
and
\begin{equation}
\label{Eq:PTliq}
\begin{split}
 P_{\text{T}}^{\text{IK}}(z) = &\rho(z)k_{\text{B}}T
   - \frac{1}{2A} \left\langle
\sum_{i<j}\frac{x_{ij}^{2}+y_{ij}^{2}}{r_{ij}} 
\frac{u^{\prime}_{\text{pp}}(r)}
{\mid
z_{ij}\mid}  \right.\\
 & \left. \times  \Theta \left(\frac{z-z_{i}}{z_{ij}}\right)
\Theta\left(\frac{z-z_{i}}{z_{ij}}\right)\right\rangle
\end{split} ,
\end{equation}
where $\theta$ is the Heavyside step function, $z_{ij}=z_{j}-z_{i}$,
and $\rho(z)$ is the local density given by
\begin{equation}
 \rho(z)=\frac{N(z)}{(A/2) \times \Delta z} \; .
\end{equation}
Here, $\Delta z$ is the bin width used to obtain the pressure profiles
and $N(z)$ is the number of liquid particles in the bin between $z$ and
$z+\Delta z$. This contribution to the local pressure tensor is added
to all bins between $z_{i}$ and $z_{j}$.  It is to be noted that the
liquid-structured wall interaction has no contribution to the tangential
component of the pressure tensor due to the periodicity of our system
in the lateral direction~\cite{nijmeijer90,vanleeuwen90}.

The contribution to the pressure tensor from the structureless walls
can also be taken into account by the IK 
method~\cite{varnik00,varnik-thesis00}
by considering  the  walls 
at $z_{\text{b}}$ and $z_{\text{t}}$ to be particles of
infinite mass. From Eq.~(\ref{Eq:PNliq}), we thus obtain
\begin{equation}
\begin{split}
 P_{\text{N}}^{\text{fw}}(z)=&\frac{1}{A}\left\langle \sum_{i=1}^{N}
F_{\text{fw}}(z_{i}-z_{\text{b}})\Theta(z_{i}-z)\right\rangle- \\
 & \frac{1}{A} \left\langle\sum_{i=1}^{N}
F_{\text{fw}}(z_{\text{t}}-z_{i})\Theta(z-z_{i})\right\rangle ,
\label{eq:Pwall}
\end{split}
\end{equation}
with $F_{\text{fw}}(z)=-dU_{\text{fw}}(z)/dz$.

From Eqs.~(\ref{Eq:PNliq}) and ~(\ref{Eq:PTliq}), it is clear that
if two particles in a bin are located on the same side of $z$, their
contribution to the local pressure tensor cannot be taken into account
by the IK method. To minimize the number of such cases, we must choose
the bin width to be  comparable to the shortest distance between between
the particles in the $z$ direction. On the other hand if the bin width is
too small, there will be larger fluctuations in the pressure tensor and
the average must be taken over many more configurations to get a smooth
profile, thus increasing the computational time. In our simulations we
choose a bin-width of $\Delta z= 0.05$.

Equation~(\ref{eq:surften}) being the difference between two similar
numerical values is subject to large relative errors.  Moreover, at
large densities near the wall, the density and pressure profiles show
rapid oscillations and hence resolving them with high precision requires
a huge computational effort. Below, the accuracy of the PA method is
studied in detail via a direct comparison to the data obtained from
TI.

\subsection{$\gamma$ from TI}
In a TI, the free energy of a state of interest is
computed with respect to a reference state~\cite{frenkel-smit02}.  A parameter
$\lambda$, which couples to the interaction potential, is gradually
changed such that the reference state is reversibly transformed into
the final state of interest.

To calculate the interfacial free energy of the LJ system in contact
with a structured wall, the TI scheme is carried
out in two steps. In the first step, a bulk LJ system without walls
and periodic boundary conditions in all directions is reversibly
transformed into a LJ system in contact with a structureless flat wall
along the $z$ direction.  In the second step, the flat wall interacting
with the LJ system is reversibly transformed into a structured wall.
To ensure reversibility of the thermodynamic path, periodic boundary
conditions are applied in $x$, $y$ and $z$ direction. Calculating the
free energy change in the two steps yields the required interfacial
free energy.

\begin{figure}
\includegraphics[width=3.0in]{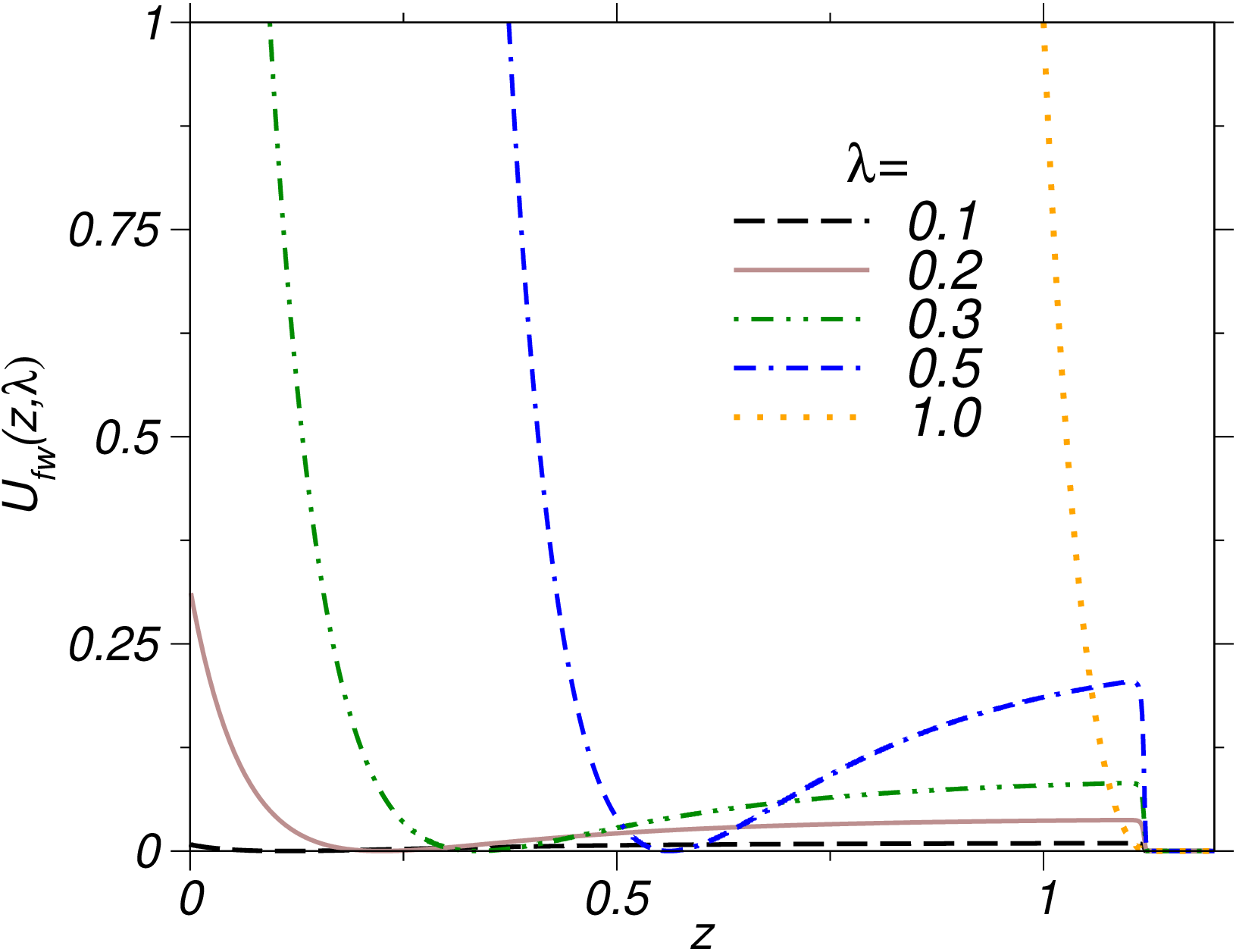}
\caption{\label{fig:1}(Color online)  
Variation of the WCA wall potential as a function of $\lambda$ during
the transformation of the bulk LJ liquid or crystal into an impenetrable
flat wall interacting with the LJ system.  From no wall at $\lambda=0$,
we have a wall with a finite barrier at small values of $\lambda$. With
increasing $\lambda$, the wall becomes more and more impenetrable.
At $\lambda=1$, there is an impenetrable wall represented by the WCA
potential.}
\end{figure}
To obtain $\gamma$ for a hard-sphere system in contact with a
hard structureless wall via TI, Heni and
L{\"o}wen~\cite{heni99} have used a scheme, where a bulk hard sphere
system is reversibly transformed into a system interacting with an
impenetrable hard wall. In this work, we generalize the scheme of Heni
and L{\"o}wen to continuous wall potentials. To this end, the wall
potential is parametrized by a parameter $\lambda$ such that the wall
changes smoothly from a penetrable to an impenetrable wall as $\lambda$
increases. The following parametrization of the wall potential is adopted:
\begin{equation}
\begin{split}
\label{eq:uwallsubst}
u_{\text{fw}}(\lambda,z)=&
\lambda^{2}4\epsilon_{\text{w}}\left[\left(\frac{\sigma_{\text{pp}}}{
z+(1-\lambda)z_{\text{cw}}} \right)^{12}\right.
\\ & \left.-\left(\frac{\sigma_{\text{pp}}}{z+(1-\lambda)z_{\text{cw}}}
\right)^{6}+\frac{1}{4}\right] \times w(z) .
\end{split}
\end{equation}
Figure~\ref{fig:1}, shows the parametrized wall potential at
different values of $\lambda$.  At $\lambda=0$ a bulk LJ system can freely
cross the boundaries. For small values of $\lambda$ the barrier height
at $z=0$ is of the same order as $k_{\text{B}}T$ and the LJ particles
can penetrate the barrier. As $\lambda$ increases, the wall becomes more
and more impenetrable and finally an impenetrable WCA wall is obtained
at $\lambda=1$.

Since, the interfacial excess free energy of the LJ system in presence of
walls is calculated with respect to a bulk LJ crystal, it is important
that the bulk density is maintained as the parameter $\lambda$
is varied during the transformation. This is particularly important
for a LJ liquid close to coexistence, since an increase in the bulk
density in the presence of walls could lead to a precrystallization of
the bulk liquid during the transformation, thus making it irreversible.
A constant bulk density also ensures that our system is large enough such
that there are no mutual influences between the walls on either side of
the bulk LJ system.  To maintain a constant bulk density one must keep
the normal pressure $P_{\text{N}}$ constant and change the volume, as
is achieved by carrying out simulation in the $NP_{\text{N}}AT$ ensemble.

The system Hamiltonian now depends on $\lambda$ and is given by
\begin{equation}
\begin{split}
H({\bf r},{\bf p},\lambda)=  
\sum_{i=1}^{N_{\text{p}}} \frac{1}{2m_{i}}{{\bf p}_{i}^{2}}+
\sum_{i=1}^{N_{\text{p}}}\sum_{j=i+1}^{N_{p}}  u_{\text{pp}}({r_{ij}}) +\\
\sum_{i=1}^{N_{\text{p}}}u_{\text{fw}}(\lambda,z=z_{i}-Z)
\end{split} .
\label{eq:hamilt_lambda}
\end{equation}
and thus the partition function can be written as 
\begin{equation}
\begin{split}
Q(\lambda)=\frac{1}{h^{3N}N!}\int \int \int \exp \left
[-\frac{H({\bf r},{\bf p},\lambda)
+P_{\text{z}}AL_{\text{z}}}{k_{\text{B}}T}\right] \\
\times A dL_{\text{z}} d\textbf{p}^{N} d\textbf{p}^{N}.
\end{split}
\end{equation}

The derivative of the Gibbs free energy with respect to $\lambda$ is
\begin{equation}
 \frac{\partial G(\lambda)}{\partial \lambda} =
-\frac{k_{\text{B}}T}{Q(\lambda)}
\left[\frac{\partial Q(\lambda)}{\partial \lambda}\right]
= \left \langle \frac{\partial H(\lambda)}{\partial\lambda} 
\right\rangle_{\lambda}\,
\end{equation}
where the angular brackets denote the ensemble average at a particular
value of $\lambda$ in the $NP_{\text{N}}AT$ ensemble.

The Gibbs free energy difference between the two initial and final
state can then be obtained as
\begin{align}
\label{eq:DG}
\Delta G &= G(\lambda=1)-G(\lambda=0)=\int_{0}^{1} 
\left [ \frac{\partial
G(\lambda)}{\partial \lambda} \right ]_{\lambda} d\lambda \\
\label{eq:delGint}
&=\int_{0}^1 \left \langle 
\frac{\partial H(\lambda)}{ \partial \lambda} \right
\rangle_{\lambda} d\lambda \; .
\end{align}
To compute $\Delta G$ from molecular simulations, independent
simulations runs are carried out at $N_{\lambda}$ discrete intervals
between $\lambda=0$ and $\lambda=1$.  Alternatively, one can also
calculate the free energy difference in a single simulation by varying
$\lambda$ step by step such that the final configuration at a value
of $\lambda=\lambda_{i}$ is the initial configuration for the next
value at $\lambda=\lambda_{i+1}$. In both methods, the system is
equilibrated at each $\lambda=\lambda_{i}$, and then the time average
of the quantity $\partial H(\lambda)/\partial \lambda$ is calculated.
The numerical integration of Eq.~(\ref{eq:delGint}) is carried out using
the trapezoidal rule:
\begin{equation}
 \Delta G = \sum_{i=1}^{N_{\lambda}-1} \frac{1}{2} 
\left[\left\langle \partial H /\partial \lambda \right \rangle_{i} +
\left\langle \partial H /\partial \lambda \right \rangle_{i+1}\right ]
(\lambda_{i+1}-\lambda_{i}) \; .
\label{eq:DGTI}
\end{equation}
The partial derivative of $H(\lambda)$ with respect to $\lambda$ is
given by
\begin{equation}
\label{eq:derH}
\begin{split}
\frac{\partial H(\lambda)}{\partial \lambda}=\frac{\partial
u_{\text{fw}}(z,\lambda)}{\partial \lambda}=
\frac{2}{\lambda}u_{\text{fw}}(\lambda,z)+ \\
\frac{z_{\text{cw}}}{z+(1-\lambda)z_{\text{cw}}}u_{\text{fw}}(z,\lambda) \; .
\end{split}
\end{equation}

The above TI scheme leads to a wall which is not
fully impenetrable and hence does not correspond to the desired state
of interest at the end of the integration path. While the LJ particles
cannot cross the wall at $\lambda=1$, two particles near the boundary
but on opposite sides of the wall can still interact with each other. To
overcome this problem another TI step is carried
out to bring the system to a state where the LJ particles are in contact
with a fully impenetrable wall excluding such spurious interactions. This
is achieved by parametrizing $u_{\text{pp}}(r)$ by a factor $\mu$,
\begin{equation}
\begin{split}
u_{\text{pp}}(\mu,r_{\text{ij}})= {u_{\text{pp}}^{(1)}(r_{ij})}+
(1-\mu){u_{\text{pp}}^{(2)}(r_{ij})}
\end{split}
\end{equation}
where, ${u_{\text{pp}}^{(1)}(r_{\text{ij}})}$ denotes
interaction between LJ particles on same side of the wall, while
${U_{\text{pp}}^{(2)}(r_{ij})}$  corresponds to interaction between
particles near the  boundary but on opposite sides of the wall, i.e.~the
separation between particles is greater than $L_{z}/2$. At $\mu=0$, all
such spurious interactions are taken into account.  As $\mu$ increases such
interactions are reduced by the factor $1-\mu$ and finally at $\mu=1$,
these spurious interactions are completely neglected.  The $\mu$ dependent
Hamiltonian for this step can be written as,
\begin{equation}
\begin{split}
H({\bf r},{\bf p},\mu)=  
\sum_{i=1}^{N_{\rm p}} \frac{1}{2m_{i}}{{\bf p}_{i}^{2}}+\\
\sum_{i=1}^{N_{\rm p}}\sum_{j=i+1}^{N_{p}}  \left [u_{\text{pp}}^{(1)}(r_{ij}) 
+ (1-\mu)u_{\text{pp}}^{(2)}(r_{ij})\right]  +\\
\sum_{i=1}^{N_{p}}u_{\text{fw}}(\lambda=1,z=z_{i}-Z) \; .
\end{split}
\label{eq:hamilt_mu}
\end{equation}

The thermodynamic integrand in this step is
\begin{equation}
\partial H /\partial \mu = \partial u_{\rm pp}(r,\mu) /\partial \mu 
= - u_{pp}^{(2)}(r_{\text{ij}}) .
\end{equation}
and thus the free energy difference can be expressed as
\begin{equation}
\Delta G_{{\rm fw}\rightarrow {\rm fw}^{\ast}}=
\int_{0}^{1}\langle \partial u_{\rm pp}(r,\mu) /\partial \mu \rangle_{\mu} d\mu .
\end{equation}
Our simulations showed that the contribution of $\Delta G_{{\rm
fw}\rightarrow {\rm fw}^{\ast}}$ is very minor, i.e.~about 0.1\% of
$\Delta G$ from Eq.~(\ref{eq:DGTI}) and hence can be neglected.

Using Eqs.~(\ref{eq:defife}), (\ref{eq:DG}), and (\ref{eq:derH}), the
interfacial free energy of a LJ system interacting with a flat wall can
be written as
\begin{align}
 \gamma=\frac{G_{\text{fw}}-G_{\text{bulk}}}{A}=
\frac{{\Delta G}_{\text{bulk} \rightarrow \text{fw}}}{A}
\end{align}
with
\begin{equation}
 {\Delta G}_{\text{bulk}\rightarrow \text{fw}}=\int_{0}^{1}\left \langle
\frac{\partial u_{\text{fw}}(z,\lambda)}{\partial \lambda}
\right\rangle_{\lambda} d\lambda .
\end{equation}

In the second step of our TI scheme, the flat
wall is reversibly transformed into a structured wall in contact with
the LJ system.  During this change, the flat walls are positioned at the
same location as the structured wall layer closest to the LJ liquid or
crystal and there is no interaction between the flat and structured walls.
The transformation from flat walls to structured walls is accomplished
by parametrizing the wall potential as:
\begin{equation}
\label{eq:uwall}
U_{\rm wall}(r_{\text{ij}},\lambda)= 
(1-\lambda)^{2} \sum_{i}u_{\text{fw}}(z=z_{i}-Z) + \\ 
\lambda^{2} \sum_{i,j}u_{\text{pw}}({r_{\text{ij}}}) \; .
\end{equation}
Now, the $\lambda$-dependent Hamiltonian is
\begin{equation}
\begin{split}
H({\bf r},{\bf p},\lambda)=  
\sum_{\text{i=1}}^{N_{\rm p}} \frac{1}{2m_{i}}{{\bf p}_{\text{i}}^{2}}+
\sum_{\text{i=1}}^{N_{\rm p}}\sum_{\text{j=i+1}}^{N_{p}}  
u_{\text{pp}}(r_{\text{ij}}) +\\ 
(1-\lambda^{2})\sum_{i=1}^{N_{\rm p}}u_{\text{fw}}(z=z_{i}-Z,\lambda)+
\lambda^{2} \sum_{i=1}^{N_{\rm p}}\sum_{j=1}^{N_{\rm w}}u_{\text{pw}}(r_{\text{ij}})
\end{split}
\label{eq:hamilt_swfw}
\end{equation}
and the derivative of the Hamiltonian with respect to $\lambda$
\begin{equation}
\begin{split}
\frac{\partial H}{\partial \lambda} = & 
\frac{\partial U_{\text{wall}}}{\partial \lambda}\\
=& 2\left [(\lambda-1)\sum_{i}u_{\text{fw}}(z=z_{i}-Z)+
\lambda \sum_{i=1}^{N_{\rm p}}\sum_{j=1}^{N_{\rm w}} 
u_{\text{pw}}(r_{\text{ij}})\right] \; .
\end{split}
\end{equation}
So, finally the interfacial free energy of the LJ system in contact 
with a structured wall (sw) is given by
\begin{align}
\gamma_{\text{wc}}=\frac{G_{\text{sw}}-G_{\text{bulk}}}{A}=
\frac{{\Delta G}_{\text{bulk} \rightarrow \text{fw}}}{A} 
+\frac{{\Delta G}_{\text{fw}\rightarrow \text{sw}}}{A}
\end{align}
with
\begin{equation}
{\Delta G}_{\text{fw}\rightarrow \text{sw}}=
\int_{0}^{1}\left \langle
\frac{\partial U_{\text{wall}}(\lambda)}{\partial \lambda}
\right\rangle_{\lambda} d\lambda \, .
\end{equation}
\subsection{Simulations}
To integrate the equations of motion, the velocity form of the Verlet
algorithm was used with a time step $\tau=0.005$ and, to maintain constant
normal pressure, the Andersen barostat algorithm~\cite{andersen80}
was chosen. Periodic boundary conditions are employed in the $x$,
$y$ and $z$ directions for the first step of the TI method where flat
walls are considered. In the second step periodic boundary conditions
are only used along the $x$ and $y$ directions.
The PA simulations are carried out with periodic boundary conditions only
along the $x$ and $y$ directions.  The temperature was kept constant
by drawing every 200 steps the velocity of the LJ particles from the
Maxwell-Boltzmann distribution at the desired temperature.

During the $NP_{\rm N}AT$ simulations, the position of the flat or structured
walls must be modified keeping the normal pressure $P_{\text{N}}$
constant.  To ensure this, the flat walls are treated as particles
of infinite mass and, at each time step, the wall position $z_{\text{fw}}$
is rescaled according to
\begin{equation}
Z(t+\Delta t)=Z(t)\times L_{\text{z}}(t+\Delta t)/L_{\text{z}}(t) \; .
\label{eq:fwpconst}
\end{equation}
Note that this method is similar to the ``fluctuating wall" method 
proposed by Lupowski and van Swol~\cite{lupkowski90}, 
maintaining a constant normal pressure in a MC simulation of 
LJ particles in presence of a structureless wall.

When a rigid structured wall interacts with the LJ system, the wall
particles must not change their positions relative to each other,
thereby changing the wall density. To circumvent this problem, the
center of mass of the wall is changed at every time step according to
Eq.~(\ref{eq:fwpconst}).  The position of the individual particles of
the wall is then shifted such that they are at the same relative distance
from the center of mass as at the beginning of the simulation.

To calculate $\gamma_{\rm wl}$ we consider systems of $4000$ particles.
The structured walls contain between 200-1200 particles, depending on
the orientation and the density of the wall.  The total surface area
of the simulation cell is about $A=200$, yielding a length along the
$z$ direction of about $L_{\text{z}}=65$ at the various wall-liquid
interaction strengths $\epsilon_{\rm pw}$ and structured wall densities
$\rho_{\text{w}}$. $\gamma_{\rm wl}$ is computed at a normal pressure of
$P_{\rm N}=3$ and temperature $T=2$.  At the start of the simulation,
the LJ particles were placed on ideal fcc lattice sites and the walls
were inserted simultaneously. Then the system was allowed to melt
and equilibrate at the desired pressure and temperature, before the
calculations were performed.

To test for the presence of any finite size effects, we also performed
simulations with up to 12000 particles and a total surface area of about
$A=340$, but obtained identical results compared to the simulations
carried out with the smaller system size. This shows that systems of
$4000$ particles are large enough to avoid finite size effects in the
calculation of interfacial free energies.

From previous works pertaining to hard sphere systems, it is well known
that the (111) orientation of the crystal in contact with a planar hard
wall (or a soft WCA wall) gives the lowest interfacial tension as compared
to the (100) or (110) orientations~\cite{courtemanche93}. At small 
undercoolings, the hard sphere fluid freezes into the (111) crystal 
near the wall~\cite{dijkstra04}.
Hence, we obtain interfacial free energies only for the (111) orientation
of the fcc crystal phase in contact with the walls.  Unlike the liquid,
the crystal has a long-range order and, in order to prevent deformation
of the crystal, the system size must be compatible with this order.

For the determination of $\gamma_{\rm wc}$, systems of 7056 particles
and area around $A=450$ are considered.  The number of structured wall
particles ranges from $800$ to about $1200$, depending on the different
wall densities.  Only the (111) orientation of the crystal in contact with
the (111) orientation of the structured wall along the interface normal
was considered.  The corresponding simulations to obtain the interfacial
free energy of a crystal in contact with a flat wall are carried out with
$3960$ particles with an area of around $A=200$. Simulations were also
carried out with a system size of $6006$ particles and a total area of
$A=300$ and there was only a marginal deviation ($<1\%$) in the value
of $\gamma_{\rm wc}$ as compared to the smaller system.

For comparing results obtained by our approach with that of 
Tang-Harris~\cite{tang95} and Grzelak-Errington~\cite{grzelak08}, we 
performed simulations for their system with $4000$ liquid particles 
and $392$ structured wall particles at the temperature $T=0.9$.  We 
considered a lateral system size of $10 \times 10$ and the length 
of the box along the $z$ direction was kept at $60.5134$ to obtain 
a bulk liquid density of $0.661$, the value reported by 
Tang and Harris~\cite{tang95} for their simulations. Simulations were 
performed at this fixed density in the $NVT$ ensemble. With this system
size, the finite size effects were negligible.  The liquid in contact 
with the flat wall [Eq.~(\ref{eq:uwallwca})] was used as the reference 
state to calculate $\gamma_{\rm wl}$ for the liquid in contact with 
the structured wall at the same bulk density and temperature.

\begin{figure}
\includegraphics[width=3.0in]{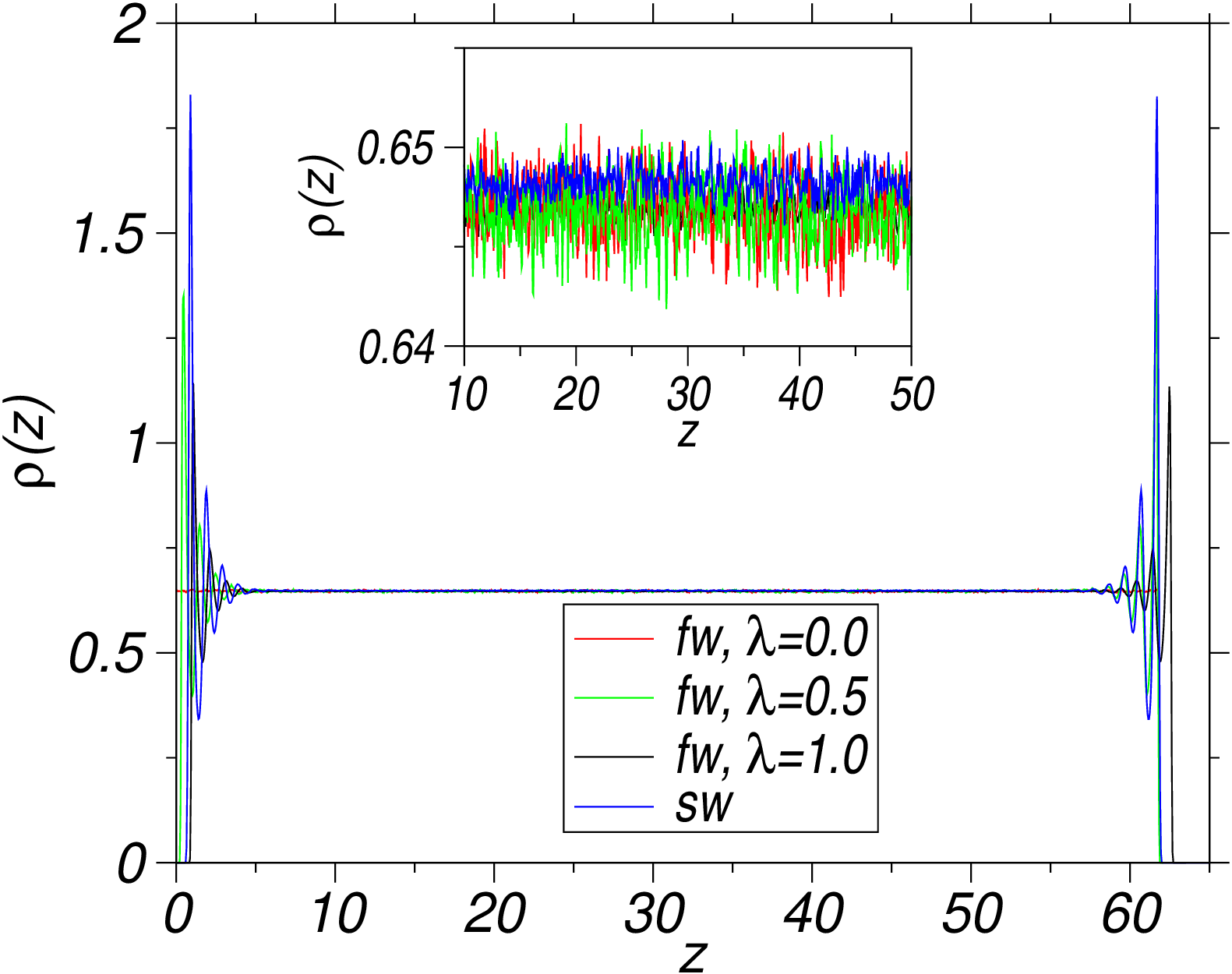}
\caption{\label{fig:2}(Color online) 
Density profile of the system configuration for different values
of $\lambda$ at the temperature $T=2.0$, the normal pressure
$P_{\rm N}=3.0$, and the wall-liquid interaction strength
$\epsilon_{\rm w}=1$. The corresponding density profile for a liquid in
contact with the (100) orientation of a structured wall of density
$\rho_{\rm w}=1.371$ at $\epsilon_{\rm pw}=1.0$ is also shown. The inset
shows the density profiles in the bulk region on a magnified scale.}
\end{figure}
\begin{figure*}
\begin{center}$
\begin{array}{cc}
\includegraphics[width=3.0in]{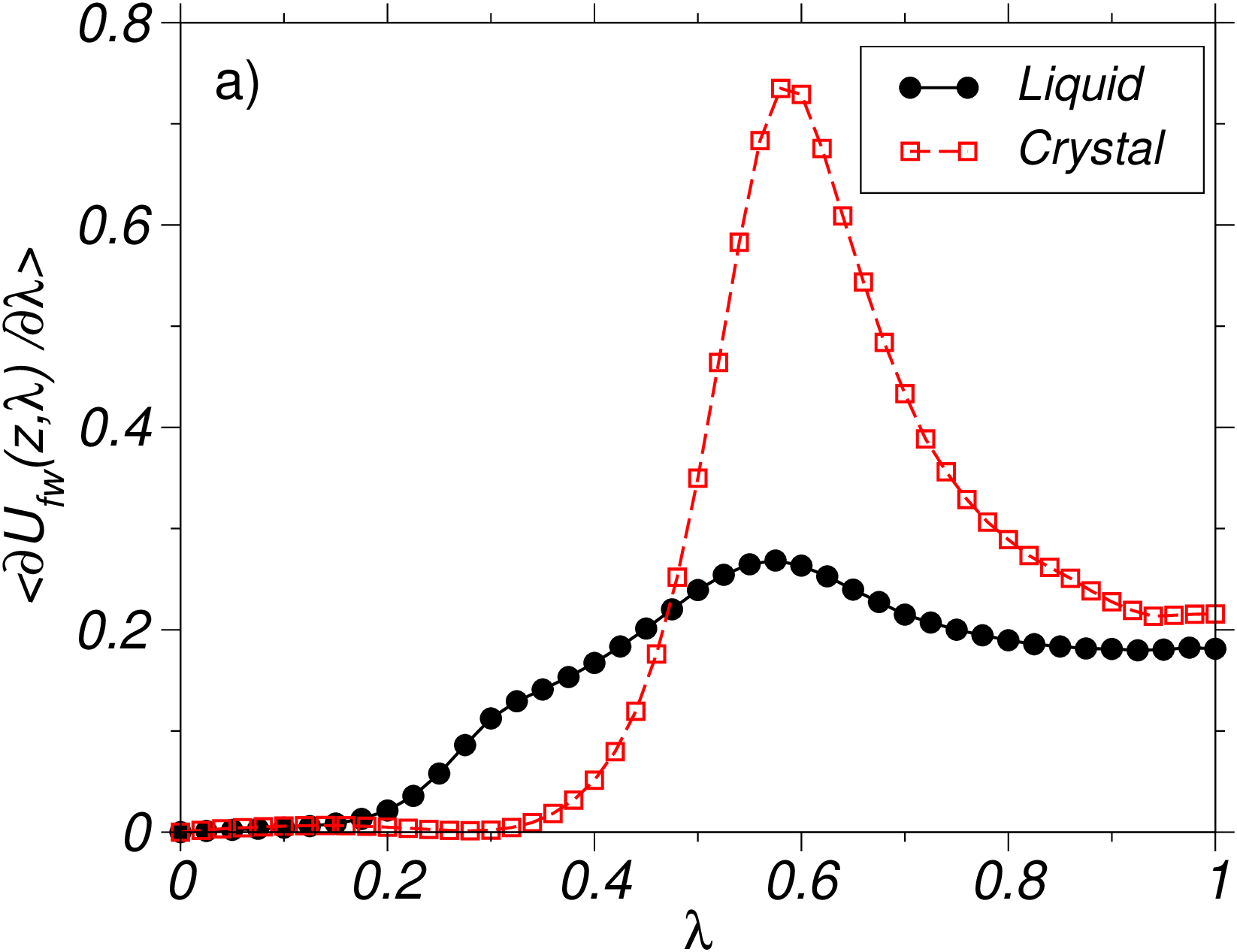} &
\includegraphics[width=3.0in]{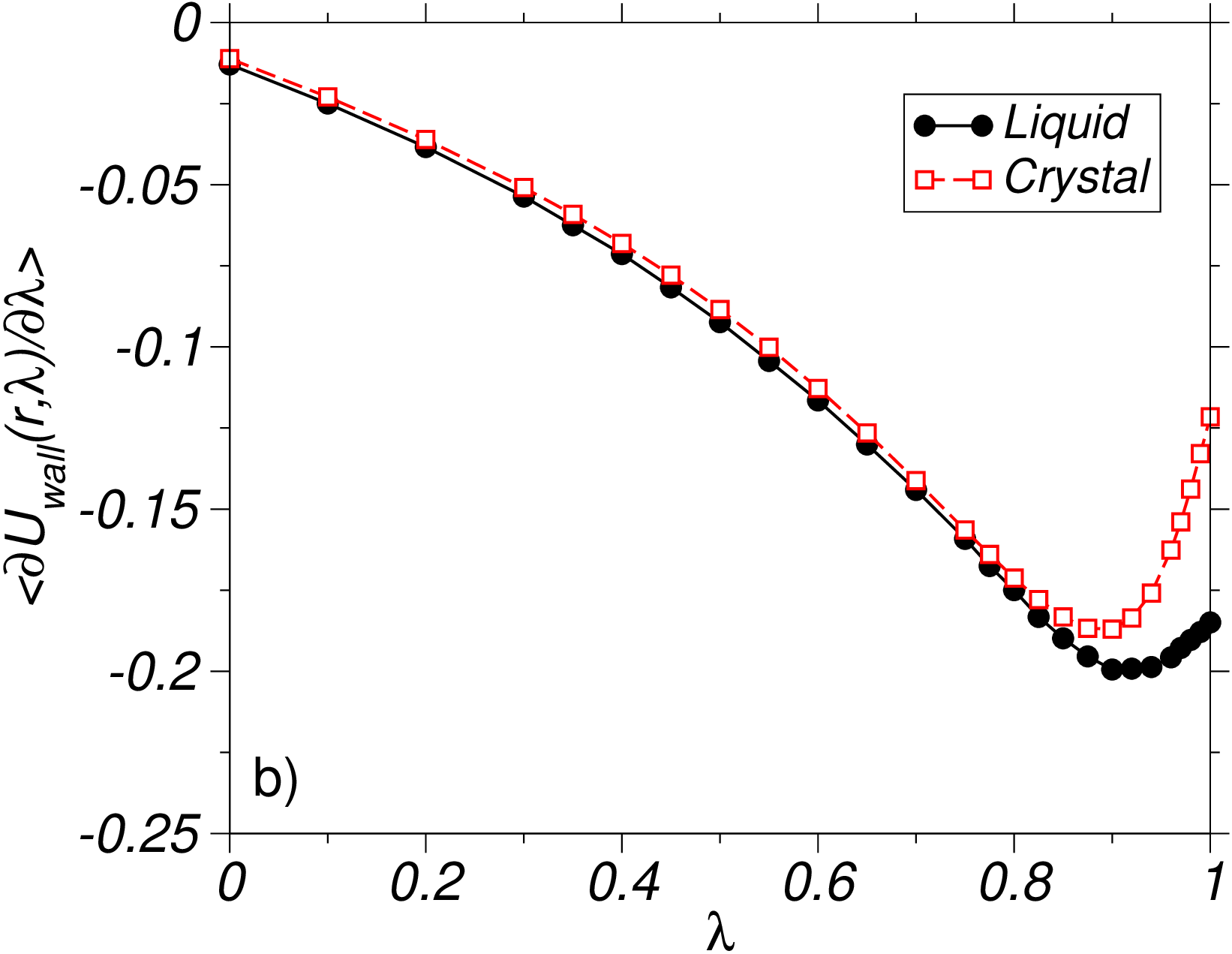} 
\end{array}$
\end{center}
\caption{\label{fig:3ab}(Color online)
(a) $\langle \partial u_{\text{fw}}(z,\lambda)/\partial \lambda\rangle$
as a function of $\lambda$ computed from simulations at $P_{\rm N}=3$
and $T=2$ for the liquid and at $P_{\rm N}=3.0$ and $T=0.5$ in case of the
crystal. To determine $\gamma_{\rm wl}$, $4000$ liquid particles 
were enclosed in a simulation box of area $A=200$ with the wall-liquid 
interaction strength $\epsilon_{\rm w}=1$. Corresponding simulations to
calculate $\gamma_{\rm wc}$ were carried out with $3960$ particles and
and area $A=235.22$. (b) $\langle \partial [U_{\rm wall}]/\partial \lambda \rangle$,
corresponding to transformation of the flat wall into  a structured
wall. In case of the liquid, the structured wall consists 
of $392$ particles rigidly fixed
to fcc lattice sites, with the (100) orientation of the wall facing
the liquid.  For the crystal, the structured wall consisted of $432$
wall particles with the (111) orientation of the wall in contact with the crystal.
The density of the structured wall was $\rho_{\rm w}=1.371$ for the 
liquid ($\rho_{\rm rm}=0.647$ for the crystal)
and the wall-particle interaction strength for the liquid -wall simulations 
was kept at $\epsilon_{\text{pw}}=1$,
while for the crystal $\epsilon_{\rm pw}=0.5$. Other parameters are same as in (a).}
\end{figure*}

\begin{figure}
\includegraphics[width=3.0in]{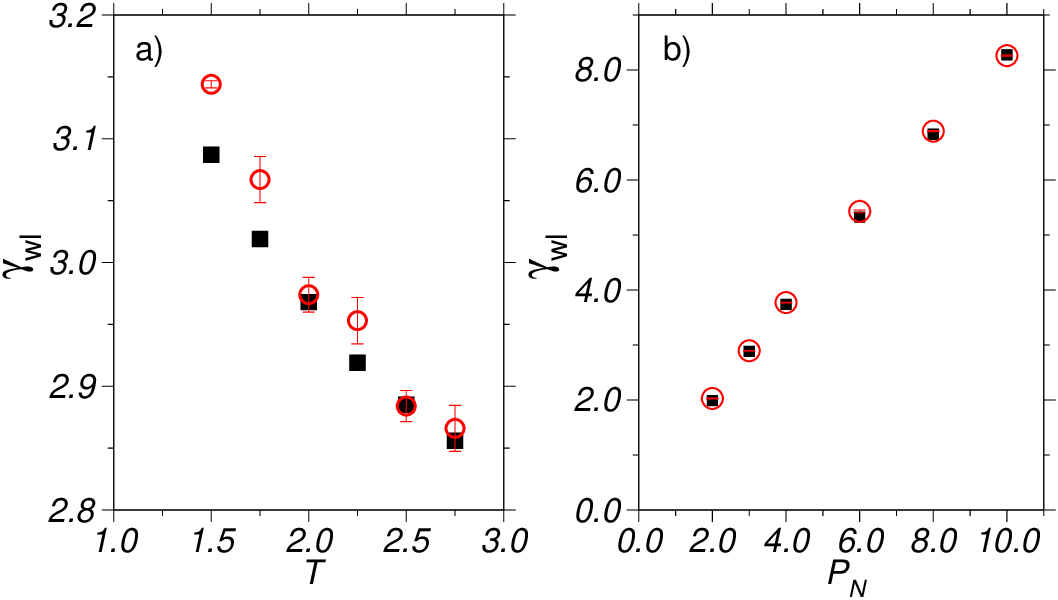}
\caption{\label{fig:4}(Color online)
(a) Interfacial free energy of liquid in contact with the flat wall,
$\gamma_{\text{wl}}$, as a function of temperature. $NP_{N}AT$ simulations
were carried out at a normal external pressure  $P_{\text{N}}=3.0$, with
$4000$ liquid particles and total surface area of $A=200$. The wall-liquid
interaction strength is $\epsilon_{\rm w}=1$. Filled squares correspond
to data obtained from TI, while estimates from PA
are represented by open circles with error bars.
Uncertainty in data computed by TI is less than
the symbol size.  (b) $\gamma_{\text{wl}}$ as a function of pressure
at $T=2.5$. Other parameters and symbols representing the TI
and PA data are same as in (a).}
\end{figure}
\begin{figure}
\includegraphics[width=3.0in]{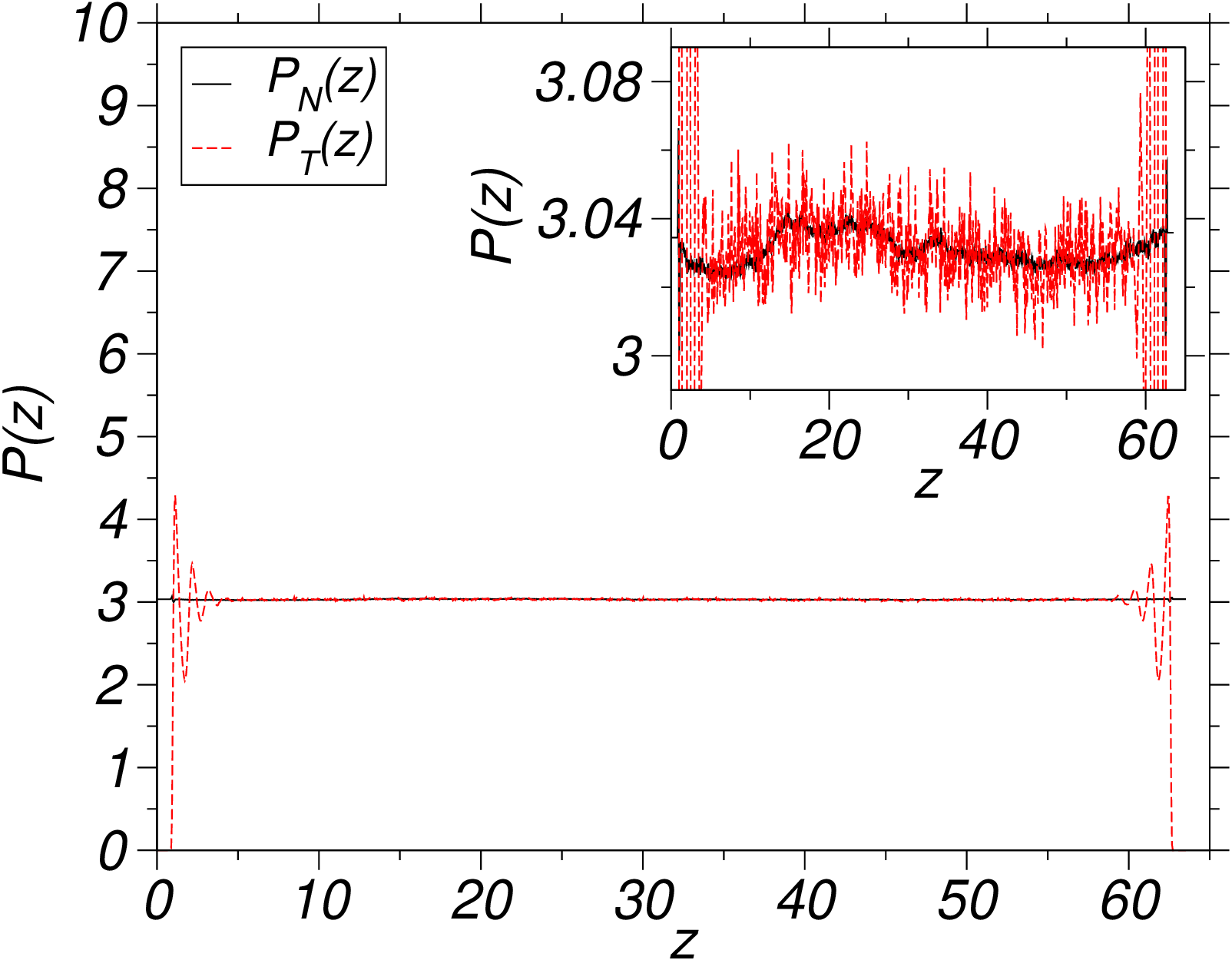}
\caption{\label{fig:5}(Color online) 
Normal and tangential components of the pressure profiles of liquid in
contact with a flat wall at $P_{\text{N}}=3$ and $T=2$, with $4000$ liquid
particles. The wall liquid interaction strength, $\epsilon_{\rm w}=1$.
Inset shows the pressure profiles on a magnified scale close to the
magnitude of the external pressure $P_{\rm N}$.}
\end{figure}

A simulation at constant normal pressure leads to fluctuations of the
length of the simulation cell in the $z$ direction, $L_{\text{z}}$.
However, in order to compute the density and pressure profiles necessary
for the PA method, it is more suitable to keep $L_{\text{z}}$ constant.
Hence, to obtain $\gamma_{\rm wl}$ via the PA method,
we first equilibrate the system in the $NP_{\rm N}AT$ ensemble
for $5\times 10^{5}$ time steps.  After equilibrium is reached, the
simulations continue for $4.5 \times 10^{6}$ time steps, from which the
average length of the box in the $z$ direction is calculated.  $L_{\rm z}$
is set to this average value and the particle coordinates are rescaled by
the factor $\langle L_{\text{z}} \rangle/L_{\text{z}}(t_{\text{f}})$,
$t_{\text{f}}$ denoting the time at the end of this equilibration
run. An equilibration run is then carried out in the $NVT$ ensemble for
$5\times 10^{5}$ time steps and the final production run consists of $4.5
\times 10^{6}$ steps, when we accumulate data for the density, energy,
temperature and pressure profiles every $5$ time steps, averaging the
profiles over $9\times 10^{5}$ sample configurations.  In our simulations,
we observe a drift of $0.5-2.5\%$ in the normal pressure profile from
the given external pressure $P_{\rm N}$.  This drift can be reduced
by averaging the length of the box for a longer simulation time or over
a large number of realizations.

To calculate the interfacial free energy via TI,
we used around $40$ intervals between $\lambda=0$ and $\lambda=1$ to
numerically compute Eq.~(\ref{eq:DGTI}).  Independent equilibration runs
were carried out at each value of $\lambda$, in the $NP_{\text{N}}AT$
ensemble for about $5\times 10^{5}-1 \times 10^{6}$ time steps.
After the completion of the equilibration run, production runs were
performed for $5\times 10^{5}$ steps in order to accumulate data. The same
TI scheme and simulation procedure has been adopted
to determine the interfacial free energy of the system investigated by
Tang-Harris~\cite{tang95} and Grzelak-Errington~\cite{grzelak08}, but in the
$NVT$ ensemble at a fixed liquid density.

For our TI method to be valid, there must be a
bulk region unaffected by the wall.  Figure~\ref{fig:2} shows
the density profile of liquid in contact with the parametrized flat wall
represented by Eq.~(\ref{eq:uwallsubst}) at various values of $\lambda$,
with the wall-liquid interaction strength $\epsilon_{\rm w}=1$.  Also
shown in Fig.~\ref{fig:2} is the density profile of the liquid
in contact with the (100) orientation of a structured wall of density
$\rho_{\rm w}=1.371$ and with interaction strength $\epsilon_{\rm pw}=1$.
The inset shows a magnified view of the density profiles in the bulk
region. Clearly, all the density profiles overlap with each other
indicating that the bulk region is unaffected by the walls.

\begin{figure}
\includegraphics[width=3.0in]{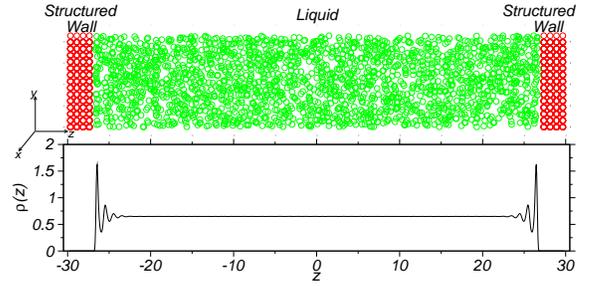}
\caption{\label{fig:6}(Color online) 
a) A two-dimensional projection onto the $zy$ plane of a sample
configuration of liquid in contact with the (100) orientation of the
structured wall at $P_{\rm N}=3$ and $T=2$. $4000$ liquid particles
are enclosed in a simulation box of area $A=200$.  The density of the
structured wall is $\rho_{\rm w}=1.371$. The wall-liquid interaction
strength is set to $\epsilon_{\text{pw}}=1$.  b) Density profile of the
liquid, averaged over many configurations, showing pronounced layering
at the structured wall-liquid interface.}
\end{figure}
\begin{figure*}
\begin{center}$
\begin{array}{cc}
\includegraphics[width=3.0in]{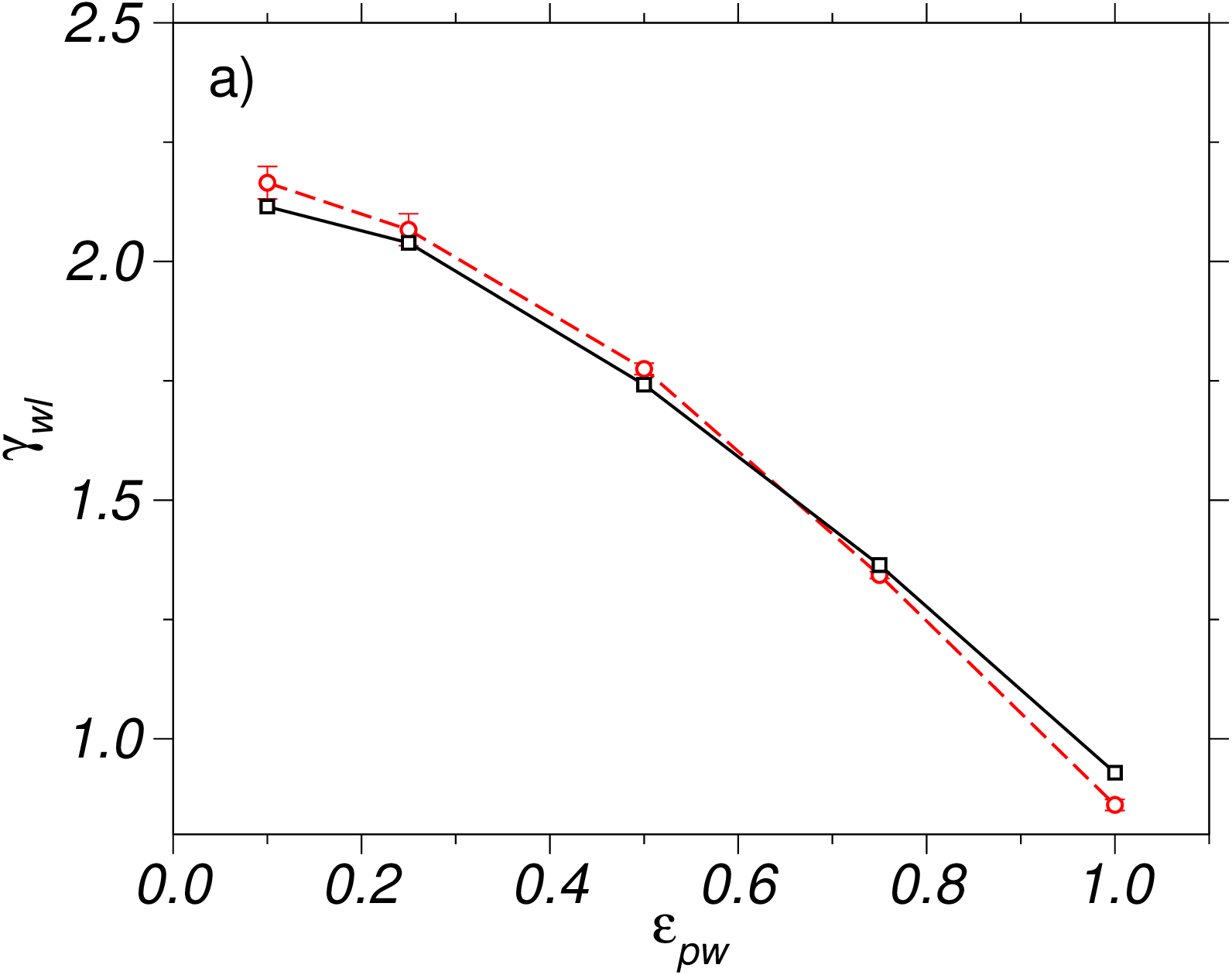} &
\includegraphics[width=3.0in]{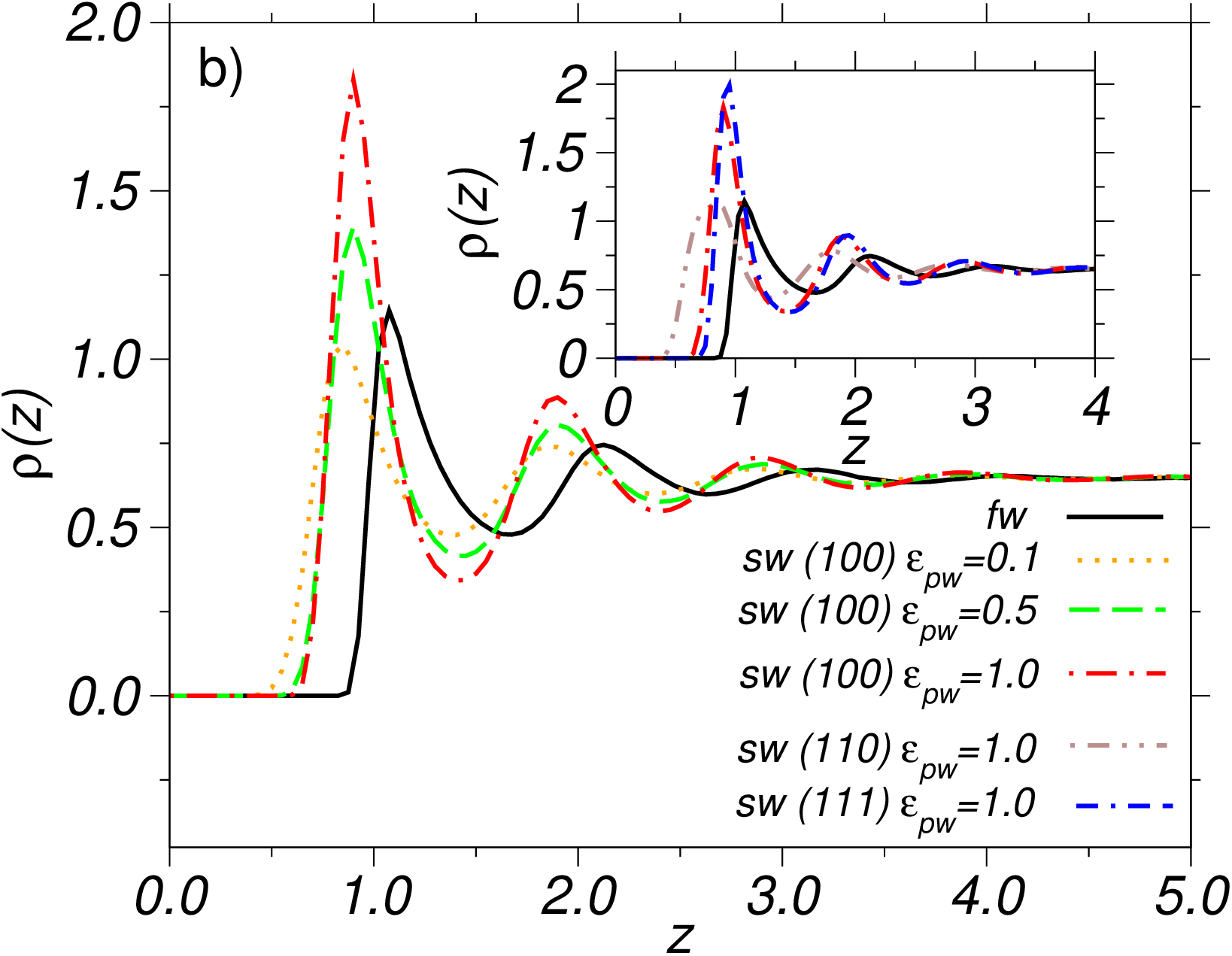}
\end{array}$
\end{center}
\caption{\label{fig:7ab}(Color online) 
a) Interfacial free energy of liquid in contact with the (100)
orientation of the structured wall vs.~$\epsilon_{\rm pw}$.
All parameters are the same as in Fig.~\ref{fig:6}.  Open squares
with solid line represent data from the TI method
and open circles with dashed line represent estimates yielded by the
PA method. Uncertainties in data from TI
are less than the symbol size. The wall-liquid interaction
strength is $\epsilon_{\rm pw}=1$.  b) Liquid density profiles
at different wall-liquid interaction strengths.  The inset shows the
profiles corresponding to the (100), (110) and (111) orientations of the
structured wall in contact with the liquid at $\epsilon_{\rm pw}=1$.
For comparison the density profile for liquid in contact with the flat
wall is also shown in both a) and b). All system parameters are same as in a).}
\end{figure*}

Figure~\ref{fig:3ab}a shows the thermodynamic integrand $\langle \partial
u_{{\text{fw}}}(z,\lambda)/\partial \lambda\rangle$ as a function of
$\lambda$, during the transformation of a bulk liquid (crystal) to a confined
liquid (crystal), interacting with flat walls. The integrand is smooth, thus
allowing for an accurate determination of the interfacial free energy.
Figure~\ref{fig:3ab}b shows the integrand as a function of $\lambda$
for the second step of the thermodynamic integration when the flat wall
is transformed into a structured wall.  The integrand is always negative,
implying that the interfacial free energy of a LJ liquid (crystal) in contact with a
rigid structured wall is smaller than for the case where the 
liquid (crystal) is in contact with a structureless flat wall.

\section{Results}
\label{sect:results}
\subsection{$\gamma_{\rm wl}$}

Using TI, we first determine the liquid-flat wall
interfacial free energy $\gamma_{\rm wl}$ at several temperatures
and pressures. In Fig.~\ref{fig:4}, we plot the data
obtained from TI along with the estimate for
$\gamma_{\rm wl}$ from the PA technique, as a function of  temperature
and pressure, respectively. The error bars in $\gamma_{\rm wl}$
obtained from TI are smaller than the symbol size
and hence are not reported.  It is evident from Fig.~\ref{fig:4}
that there is good agreement between the two methods within the
statistical error.  However, Fig.~\ref{fig:4}a shows that
$\gamma_{\rm wl}$ obtained from TI is smoothly
varying, while the PA data is less systematic. Relative differences
between the two methods are between 0.3 and 1.8\%.

This small disagreement between the two methods is due to the
large fluctuations in the local pressure profiles as obtained from
Eqs.~(\ref{Eq:PNliq}) and (\ref{Eq:PTliq}) (see Fig.~\ref{fig:5}). The
inset in Fig.~\ref{fig:5} clearly shows the large fluctuations
in the normal pressure profile near the wall and in both the
normal and tangential pressure profiles in the bulk region.  Since
Eq.~(\ref{eq:surften}) represents the difference between two pressure
profiles: Any lack of precision in the numerical measurements magnifies
the relative error.  The TI data is more accurate
and less computationally expensive compared to what would be required
to obtain more precise values from the PA method.

The liquid-flat wall system can now be used as the reference system to
calculate the interfacial free energy of the liquid in contact with a
rigid structured wall. Some properties of the structured wall such as
the wall-liquid interaction strength, density of the structured wall
and its orientation along the interface will affect the interfacial
free energy and consequently the wetting behavior of the liquid.
We will investigate $\gamma_{\rm wl}$ for the (100), (110) and,
(111) orientations of the structured wall in contact with the liquid at
different wall-liquid interaction strengths. Effects of density of the
structured wall on the interfacial free energy will also be studied.
Unless otherwise indicated, the external pressure is set to 
$P_{\rm N}=3$ and the temperature to $T=2$.

Figure \ref{fig:6} shows a sample configuration of the liquid in
contact with the (100) orientation of a structured wall in the $zy$
plane and the corresponding density profile.  We observe layering of the
particles near the wall. Away from the walls a bulk region forms, where the
density is constant.

In Fig.~\ref{fig:7ab}a, $\gamma_{\rm wl}$ is displayed as
a function of $\epsilon_{\rm pw}$, as obtained from the PA and
TI method. The error bars in the PA method were
calculated from $2-3$ realizations.  For TI,
the error bars are smaller than the size of the symbols and hence they
are not reported. We observe that $\gamma_{\rm wl}$ decreases with
$\epsilon_{\rm pw}$, which is in agreement with previous studies carried
out using the mechanical route~\cite{nijmeijer90,tang95} and other thermodynamic
methods~\cite{leroy09,grzelak08}.  High values of $\epsilon_{\rm pw}$
represent stronger attraction between the wall and liquid particles. This
reduces the free energy needed to move the liquid from the bulk to the
surface resulting in a lower interfacial free energy. Data from the two methods
agree qualitatively but the percentage difference of the PA results
with respect to the TI data  increases at higher values of $\epsilon_{\rm pw}$.

Apart from the fluctuations in the local pressure profiles, the strong
layering near the interface at large $\epsilon_{\rm pw}$ also reduces
the numerical accuracy of the PA method.  Figure~\ref{fig:7ab}b
shows the liquid density profiles at various interaction
strengths along with the corresponding profile in presence of a flat
wall. The first peak in the density profile corresponding to the
flat wall occurs at a greater distance from the wall as compared to the
peaks arising out of the liquid-structured wall interaction. This
can be attributed to the purely repulsive flat wall, which pushes the 
liquid further away from the walls compared to the structured walls.

\begin{table*}[htbp]
\centering
\begin{tabular}{|c||c|c||c|c||c|c|}
\hline
 &\multicolumn{2}{c||}{(100)}&\multicolumn{2}{c||}{(110)}&\multicolumn{2}{c|}{(111)}\\
\cline{2-7}
 $\epsilon_{\rm pw}$&$\gamma_{\rm wl}^{\rm TI}$&$\gamma_{\rm wl}^{\rm PA}$&$\gamma_{\rm wl}^{\rm TI}$&$\gamma_{\rm wl}^{\rm PA}$&$\gamma_{\rm wl}^{\rm TI}$&$\gamma_{\rm wl}^{\rm PA}$\\
\hline \hline
0.10&2.115&2.165$\pm$0.034&1.865&1.902$\pm$0.025&2.149&2.189$\pm$0.010\\
0.25&2.039&2.067$\pm$0.034&1.808&1.860$\pm$0.033&2.056&2.073$\pm$0.004\\
0.50&1.742&1.775$\pm$0.012&1.521&1.587$\pm$0.003&1.740&1.806$\pm$0.002\\
0.75&1.364&1.343$\pm$0.007&1.144&1.252$\pm$0.003&1.348&1.335$\pm$0.002\\
1.00&0.929&0.862$\pm$0.012&0.705&0.868$\pm$0.012&0.906&0.902$\pm$0.012\\
\hline
\end{tabular}
\caption{Interfacial free energy $\gamma_{\rm wl}$ at different
wall-liquid interaction strengths, for the (100), (110) and (111)
orientations of the structured wall in contact with the liquid. Data computed from
both TI and PA are shown.  Simulations are carried
out at $P_{\rm N}=3$ and $T=2$. The density of the structured wall $\rho_{\rm w}=1.371$.}
\end{table*}

The interfacial free energy of the crystal-melt
interface is influenced by the orientation of the crystal in contact
with the melt~\cite{gibbs57}. Similarly, it might be expected that 
different orientations of the structured  wall in contact with the liquid will 
affect the wall-liquid interfacial free energy.
In the inset of Fig.~\ref{fig:7ab}b, we plot the density profiles near the 
wall for the (111), (110) and (100) orientations of the structured wall in contact
with the liquid at $\epsilon_{\text{pw}}=1$. The density of the structured 
wall $\rho_{w}=1.371$ and, the lateral
dimensions of the system corresponding to the (100), (110) and (111)
orientations of the wall are $10 \times 10$, $10 \times 10.102$ and $9.623
\times 10.102$, with $392$, $420$ and $330$ wall particles, respectively.
Figure~\ref{fig:7ab}b shows the layering of the density profile to be most
pronounced for the (111) orientation owing to the closely packed 
atoms exerting a greater repulsive force on  the liquid. In contrast,
the layering for the (110) orientation is much less pronounced and the first
peak in the density profile also occurs closer to the wall as compared to the 
(111) or (100) orientations. 

In Table I, we report  $\gamma_{\text{wl}}$, obtained from the PA and
TI methods, at various $\epsilon_{\text{pw}}$, for the (100), (110) and
(111) orientations of the structured wall in contact with the liquid.
In general, we find $\gamma_{\text{wl}}$ corresponding to the (111) 
and (100) orientations of the wall to be larger compared to the (110) orientation.
This can be attributed to the stronger repulsive forces exerted on the liquid
by the more close packed (111) and (100) planes, as compared to the more loosely 
packed (110) plane. Also, relatively better agreement is 
observed between the PA and  TI methods for the (100) and (111) 
orientations of the  wall as compared to the (110) orientation.

Simulations were also carried out for large system sizes at
$\epsilon_{\rm pw}=1$, with up to $12000$ particles, and large surface area
and wall separations.  However, no systematic change was observed as
compared to the smaller system size. 
Clearly, to obtain accurate values of the interfacial free 
energy, the 
pressure profiles need to be determined with far greater numerical 
accuracy. This has also been recently pointed out by 
D.~Deb {\textit{et al.}}~\cite{deb11} for hard sphere systems. 
Computing the pressure profiles with high precision is computationally 
expensive and since accurate values can be obtained by the TI method 
with much less computational effort, use of the PA 
technique seems to be unjustified. In the remainder of the discussion on our model, 
we report results obtained with the TI method only.

\begin{figure*}
\begin{center}$
\begin{array}{cc}
\includegraphics[width=3.0in]{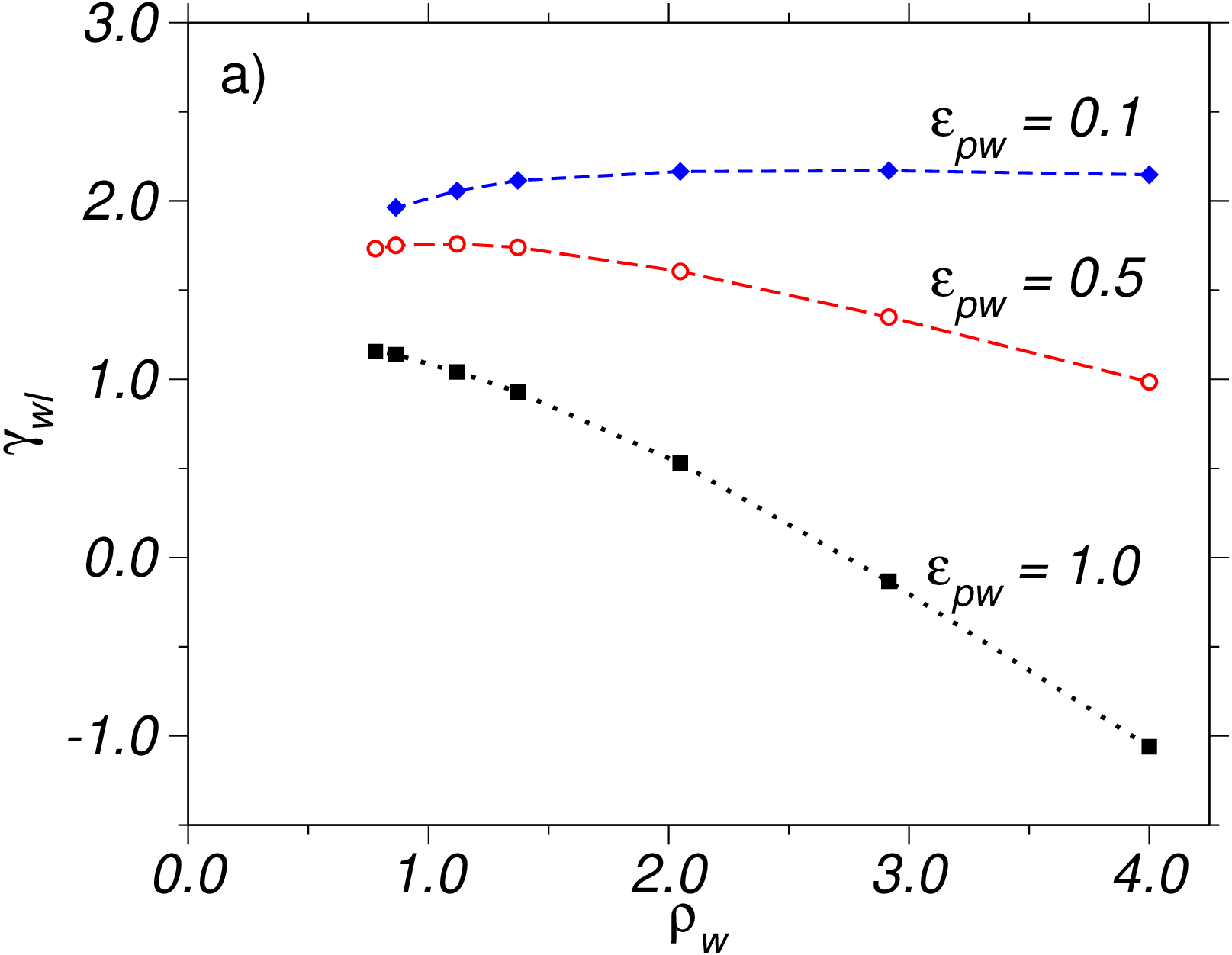}  & 
\includegraphics[width=3.0in]{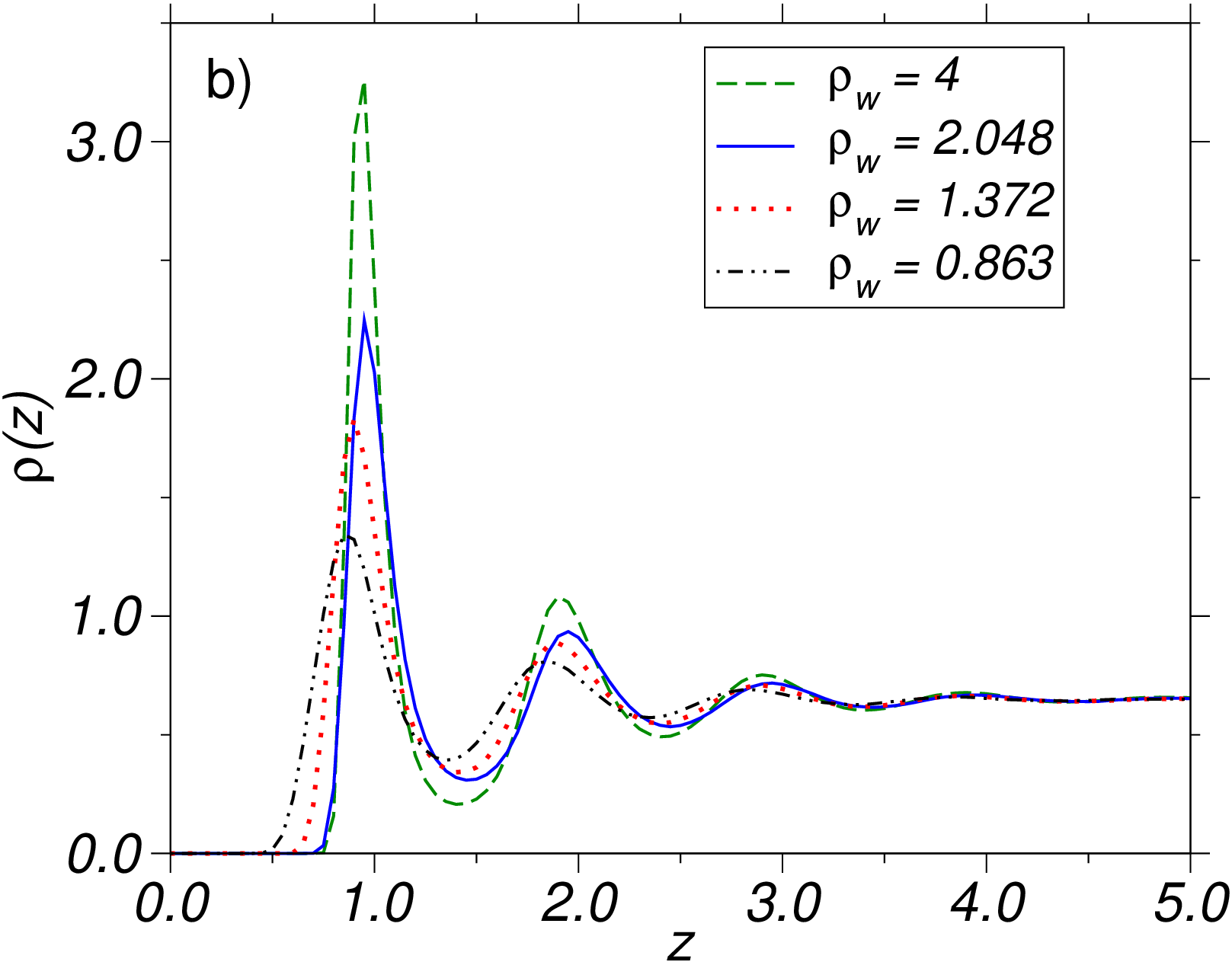}
\end{array}$
\end{center}
\caption{\label{fig:8}(Color online)
a) Interfacial free energy of a liquid in contact with a structured
wall vs.~density of the wall at several wall-liquid interaction
strengths $\epsilon_{\rm pw}$. Results were obtained from TI. 
Simulations were carried out at a constant normal pressure 
$P_{\text{N}}=3$ and temperature $T=2.0$.  b) Density profiles of the 
liquid in contact with the structured wall at various densities 
$\rho_{w}$. The wall-liquid interaction strength is $\epsilon_{\text{pw}}=1$.
Other parameters are same as in a).}
\end{figure*}
\begin{figure}
\includegraphics[width=3.0in]{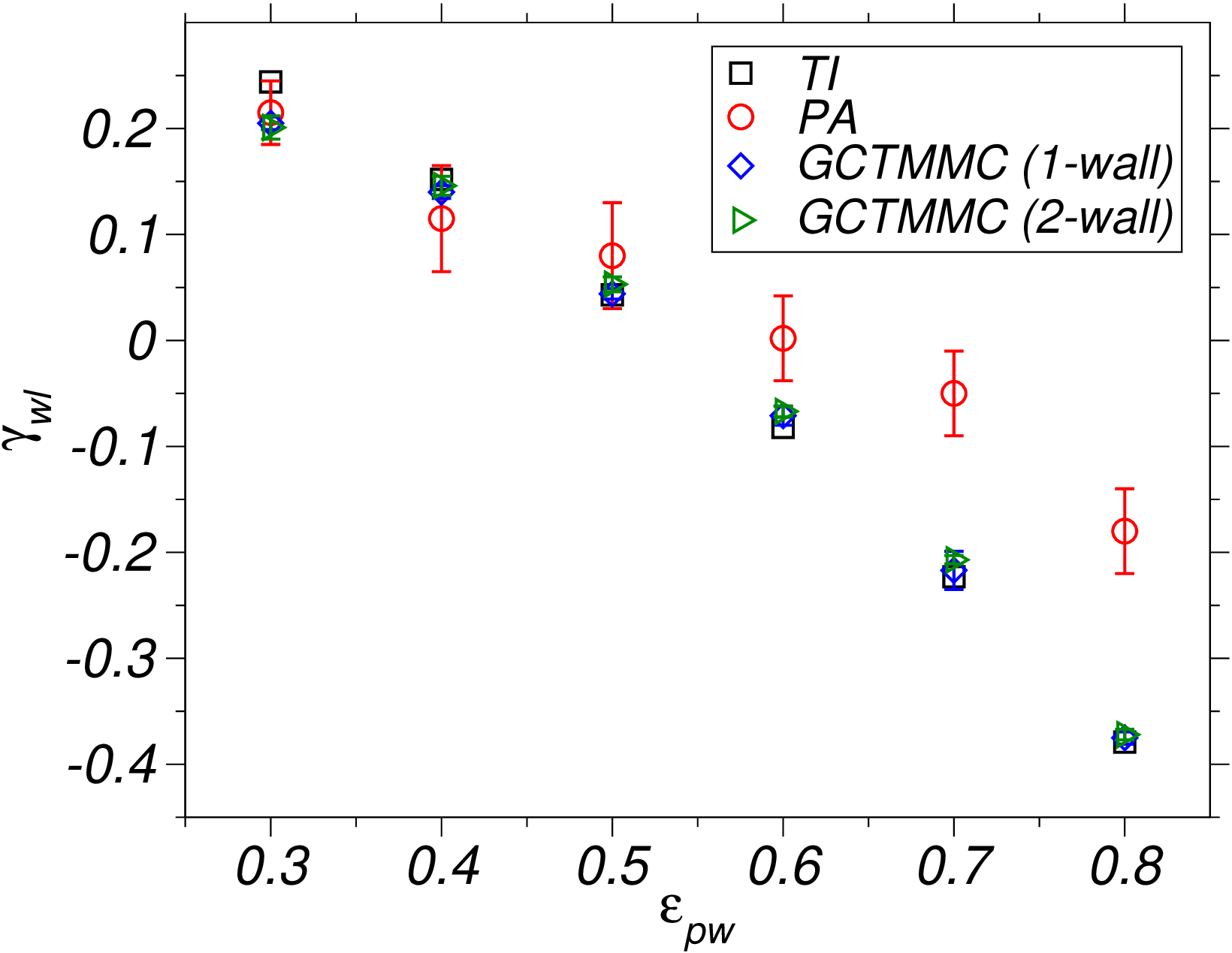} 
\caption{\label{fig:9}(Color online) 
$\gamma_{\text{wl}}$ vs.~$\epsilon_{\text{pw}}$ for the model specified
by Eq.~(\ref{eq:splj}).  Squares correspond to our results computed via
TI. Filled circles represent data from the studies of
Tang-Harris \cite{tang95}, while  diamonds and filled right triangles are the
data obtained by Grzelak-Errington \cite{grzelak08} using the one-wall and
two-wall approaches, respectively.}
 \end{figure}

The density of the structured wall will also have an impact on the wetting
behavior of the liquid in contact with it. We have carried out simulations 
at several densities $\rho_{\rm w}$ corresponding to different 
lattice constants of  the ideal fcc lattice structure of  the wall. 
In Fig.~\ref{fig:8}a, we report the TI results for $\gamma_{\rm wl}$ as a 
function of $\rho_{\rm w}$ at three different $\epsilon_{\rm pw}$'s. 
At $\epsilon_{\rm pw}=1$, the interfacial free energy decreases with the 
density of the wall. The larger number of wall particles at greater densities 
exert  strong attractive forces on the liquid, reducing the interfacial free 
energy. At extremely large densities the interfacial free energy becomes negative
indicating that the liquid  completely wets the wall.
No data for very low densities of the structured wall are shown in 
Fig.~\ref{fig:8}a since the liquid particles penetrate the wall at low densities
and the interfacial region is no longer well defined. 

At $\epsilon_{\rm pw}=0.5$,  $\gamma_{\rm wl}$ shows a weak maximum and at large
wall densities decreases with $\rho_{\rm w}$ more gradually as compared to the
situation when $\epsilon_{\rm pw}=1$. At still lower wall-liquid interaction strength
($\epsilon_{\rm pw}=0.1$), $\gamma_{\rm wl}$ has a weak dependence on the density of
the structured wall and remains almost constant in the range of $\rho_{w}$ shown in 
Fig.~\ref{fig:8}a.

In Fig.~\ref{fig:8}b, we show the density profiles of the liquid corresponding
to several densities of the structured wall $\rho_{\rm w}$ at $\epsilon_{\rm pw}=1$.
It is observed that the layering gets more pronounced and the first peak in the
density profiles occurs further away from the walls as $\rho_{\rm w}$ increases.
A similar behavior  of the density profile was observed when increasing $\epsilon_{\rm pw}$
at fixed $\rho_{\rm w}$. The variation in $\gamma_{\rm wl}$ and the nature of the density
profiles indicate that increasing $\rho_{\rm w}$ at $\epsilon_{\rm pw}=1$ has a 
similar effect on $\gamma_{\rm wl}$ as increasing $\epsilon_{\rm pw}$ at a fixed 
value of the wall density.

Finally, to compare results from our TI technique with those obtained
by other methods, we consider the model defined by Eq.~(\ref{eq:splj}),
which was first studied by Tang and Harris using a PA technique~\cite{tang95}. 
Later Grzelak and Errington~\cite{grzelak08} utilized GCTMMC simulations to obtain 
free energy profiles of the same
system over a wide range of densities, with the fluid confined
by a structured  wall on one side and a hard wall at the other side
or the fluid confined between two identical structured walls. Both of
their approaches with one  or two structured walls lead to the same results
within the statistical errors.  Using TI in the NVT ensemble, we 
computed the interfacial free energy of the same system, 
with the liquid confined by two identical structured walls.
In Fig.~\ref{fig:9}, our results are reported along with
data from the two  previous works~\cite{tang95,grzelak08}. Data obtained by
Tang-Harris  systematically deviate from our estimates of 
$\gamma_{\rm wl}$ as $\epsilon_{\rm pw}$ increases.  Their data
also has a large statistical error. However, our predictions are in 
good agreement with those of Grzelak and Errington.

\begin{figure}[htbp]
\includegraphics[width=3.0in]{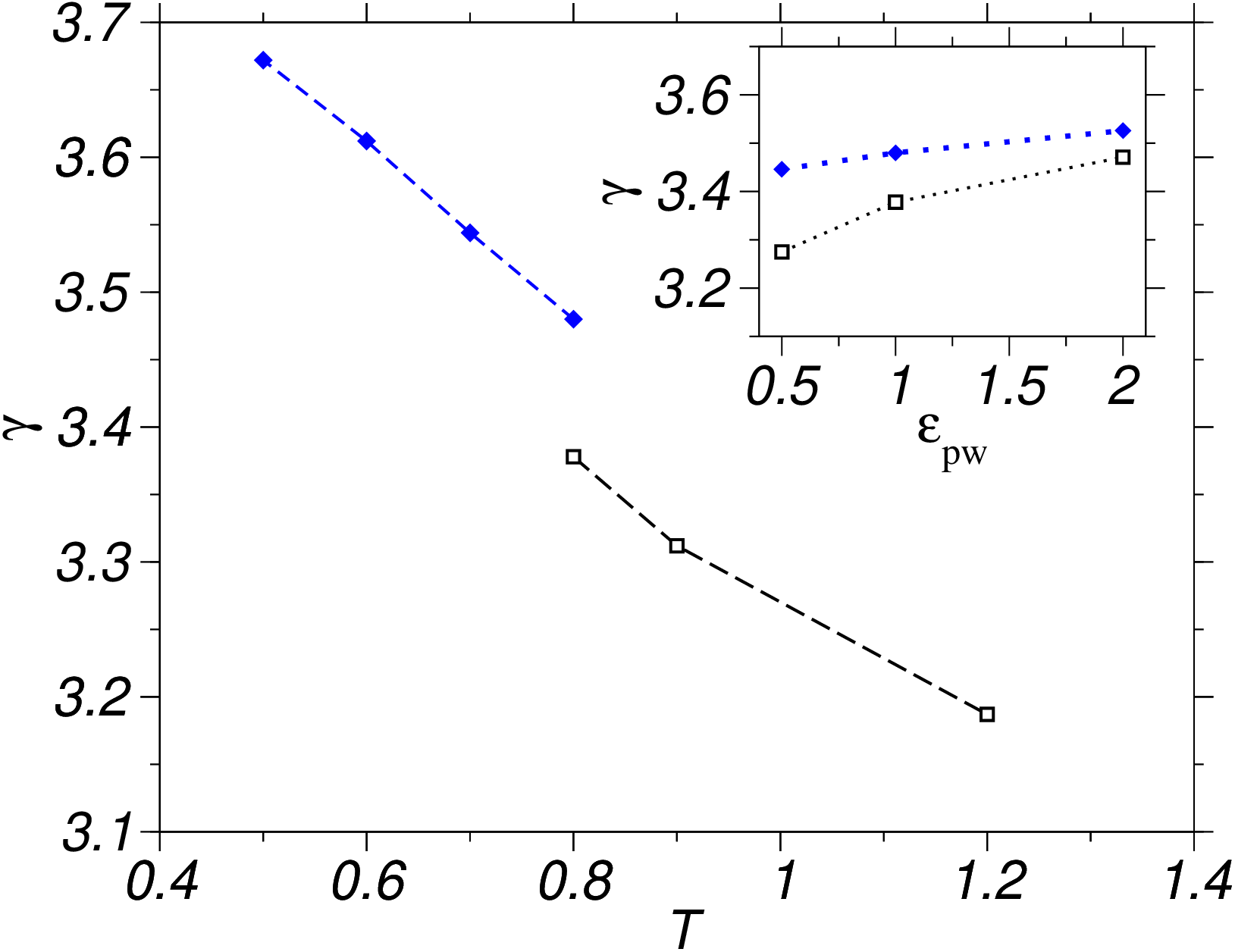}
\caption{\label{fig:10}(Color online) 
Crystal-flat wall interfacial free energy $\gamma_{\rm wl}$ (diamonds) and
liquid-flat wall interfacial excess free energy $\gamma_{\rm wc}$ (squares)
as a function of temperature. $NP_{\rm N}AT$ simulations were carried out
at $P_{\rm N}=3$ and the wall-liquid interaction strength $\epsilon_{\rm w}=1$.
The (111) orientation of the crystal was considered in determining
$\gamma_{\rm wc}$. Inset corresponds to $\gamma_{\rm wl}$
and $\gamma_{\rm wc}$ as a function of the interaction strength
$\epsilon_{\rm w}$ at coexistence: $P_{\rm N}=3.0$ and $T=0.8$. Symbols 
are same as in the main graph.}
\end{figure}

%
\subsection{$\gamma_{\text{wc}}$}
We will compute the crystal-wall interfacial energy  by the TI method
only, since the crystal can support stress and hence the interfacial
tension and interfacial free energy are not the same, thus invalidating the 
applicability  of the  PA method~\cite{tiller91}.  In performing simulations 
of crystal in contact with  walls on both sides, the number of particles $N$ 
must be chosen such that it is compatible
with the long range order of the crystal.  This is in contrast to a
liquid, where choosing a large enough $N$ yields a sufficiently large
system and the two walls on either side of the liquid do not influence each other. 
However, the crystal has a long-range order and merely choosing
a large $N$ may not necessarily be commensurate with
this order. Such an incommensurate $N$ is associated with long
range elastic distortion, that propagates from one wall to the other
 leading to an inaccurate value for $\gamma_{\text{wc}}$. This had
already been pointed out by D.~Deb {\textit{et al.}~\cite{deb10} who
studied the interfacial free energy of a hard-sphere crystal confined
between softly repulsive walls described by the WCA potential.

As specified earlier, we restrict our attention to the close-packed
(111) orientation of the crystal in contact with a flat wall along
the $z$ axis.  To evaluate $\gamma_{\text{wc}}$, a bulk fcc crystal
with the (111) orientation along $z$ axis is simulated in the $NPT$
ensemble. Periodic boundary conditions are employed  in all directions
to determine the average  equilibrium lattice constant and hence the
density of the crystal.  A fcc crystal with this density was chosen as
the initial configuration for the TI simulations to compute $\gamma_{\rm
wc}$. The length of the simulation box was chosen such that an integer
number of unit cells along the $x$, $y$ and $z$ directions adapted
exactly into the simulation box.  Then, independent simulations in
the $NP_{\text{N}}AT$ ensemble were carried out at each value  of the
$\lambda$ parameter during the two-step TI scheme.  In Fig.~\ref{fig:10},
we plot $\gamma_{\text{wc}}$ for crystal in contact with a flat wall,
as a function of temperature up to the coexistence temperature at
$P_{\rm N}=3.0$.  For comparison, $\gamma_{\rm wl}$ is also plotted
at the same pressure.  Similar to $\gamma_{\rm wl}$, $\gamma_{\rm wc}$
decreases as a function of temperature.

To predict the wetting behavior of the crystal in contact with the wall at
crystal-liquid coexistence, one needs $\gamma_{\text{cl}}$ in addition to
$\gamma_{\text{wl}}$ and $\gamma_{\text{wc}}$. However, without knowledge
of $\gamma_{\text{cl}}$, it is still possible to predict whether the
crystal will completely wet the wall ($\theta_{\text{c}}=0^\circ$)
or do so only partially  ($0<\theta_{\text{c}}<180^\circ$).
To this end, simulations were  carried out at coexistence ($P_{N}=3.0, T=0.8$) 
for the bulk liquid and crystal in contact with a flat wall at various 
interaction strengths between the wall and the bulk liquid or crystal. The 
data is reported in the inset of Fig.~\ref{fig:10}.  No wetting layer 
was observed near the walls during wall-liquid simulations, allowing for
a determination of the wall-liquid interfacial free energy directly
at coexistence.  We find that $\gamma_{\text{wc}}>\gamma_{\text{wl}}$,
showing that there is incomplete wetting of a flat wall by the (111) 
orientation of the LJ crystal and that the contact angle can be varied by changing 
$\epsilon_{\text{pw}}$.

While $\gamma_{\text{cl}}$ has not been determined in this work,
Davidchack and Laird~\cite{davidchack03,laird09},
obtained the crystal-liquid interfacial free energy at coexistence for a 
similar LJ model. The bulk liquid and crystal densities at the coexistence 
temperature $T=0.809$  for their model is same as for our system at $P_\text{N}=3.0$ 
and $T=0.8$. At $T=0.809$, they  obtained $\gamma_{\text{cl}}=0.428\pm 0.004$. 
Since the LJ model used in this work is not very different from their potential, 
it is safe to assume that the crystal-liquid interfacial free energies will 
not be far apart. Using $\gamma_{\text{cl}}=0.428\pm0.004$ and the values of 
$\gamma_{\text{wl}}$ and $\gamma_{\text{wc}}$ used in this work, we obtain contact 
angles  of $97.4^\circ$, $103.8^\circ$ and $113.6^\circ$ respectively, signifying 
partial wetting of the flat wall. This is in contrast to the hard sphere case, 
where the (111) orientation of the crystal led to complete wetting of the 
wall~\cite{deb10}. Such a situation of incomplete wetting will facilitate 
the study of heterogeneous nucleation of a crystal droplet at a wall-liquid 
boundary, and enable us to test the predictions of classical nucleation 
theory.

\begin{figure}[htbp]
\includegraphics[width=3.0in]{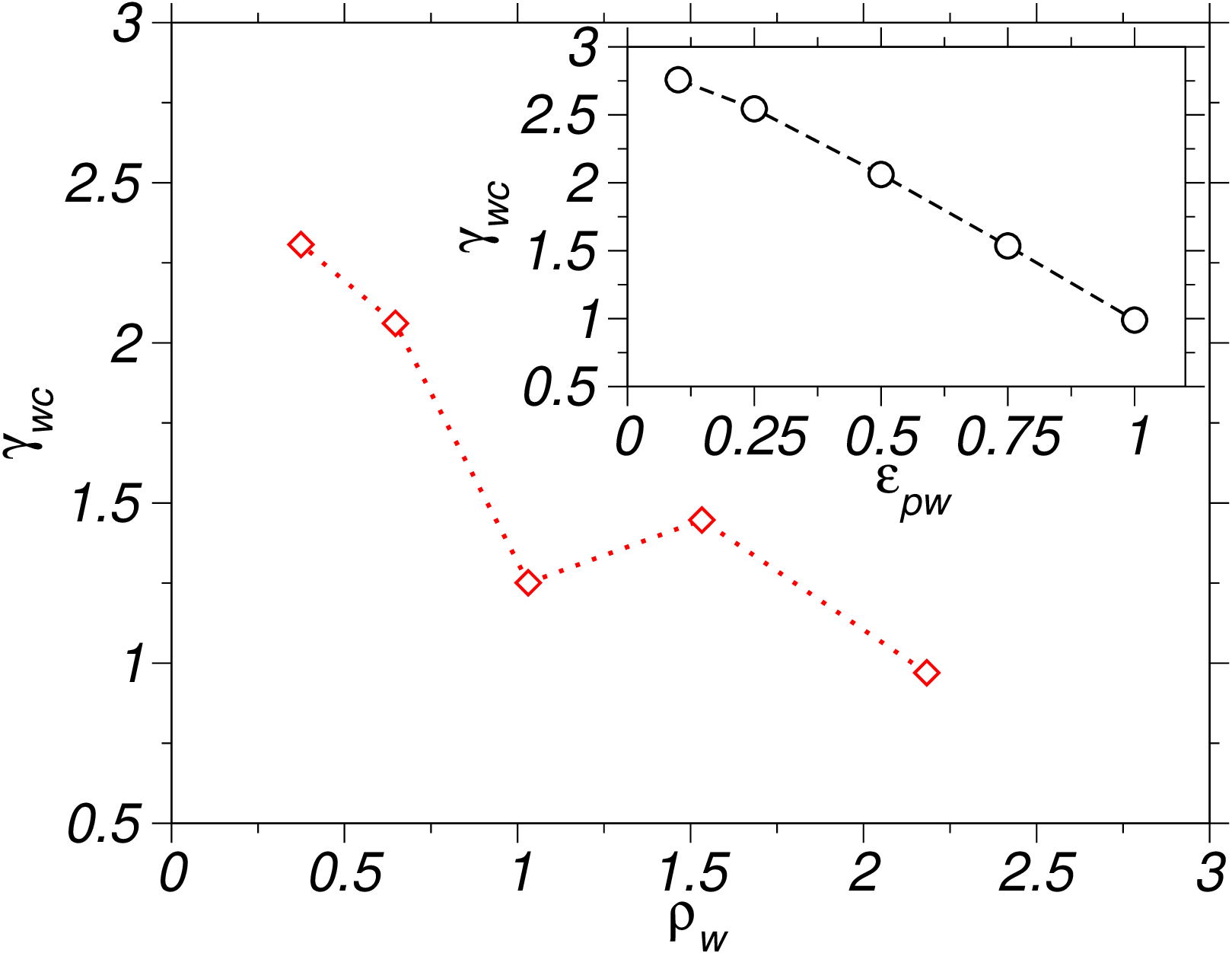}
\caption{\label{fig:11}(Color online) 
Crystal-structured wall interfacial  free energy, $\gamma_{\text{wc}}$,
as a function of the structured wall density, for the (111) orientation
of the crystal in contact with the (111) orientation of the structured
wall. $NP_{N}AT$ simulations were carried out  at a normal pressure
and temperature  $P_{\text{N}}=3$ and $T=0.5$ respectively, with the crystal-wall
interaction strength  $\epsilon_{\text{wc}}=0.5$.  The inset shows
$\gamma_{\text{wc}}$ as a function of $\epsilon_{pw}$ for a structured wall
density $\rho_{w}=0.647$, other parameters remaining the same.}
\end{figure}

Having obtained the flat wall-crystal interfacial free energy, we can
now compute the structured wall-crystal interfacial excess free energy
$\gamma_{\text{wc}}$. 
We choose to investigate  the (111) orientation of the crystal
in contact with the (111) orientation of the structured wall.
To obtain $\gamma_{\text{wc}}$, commensurate surfaces of the wall must
be  in contact with the crystal on both sides.  We know that in the fcc
structure, there is an $ABCABCABC...$ stacking of the lattice planes
along the (111) orientation. The same order of the planes must be kept
for  the crystal plane in contact with the wall. For example, the the
following stacking of the planes,
\begin{equation}
\begin{split}
A_{\text{w}}B_{\text{w}}C_{\text{w}}A_{\text{w}}B_{\text{w}}C_{\text{w}}A_{\text{c}}B_{\text{c}}C_{\text{c}}A_{\text{c}}B_{\text{c}}C_{\text{c}}...\\
......A_{\text{c}}B_{\text{c}}C_{\text{c}}A_{\text{w}}B_{\text{w}}C_{\text{w}}A_{\text{w}}B_{\text{w}}C_{\text{w}}.
\end{split}
\end{equation}
is commensurate. However, a stacking of the planes in an incommensurate
manner such as
\begin{equation}
\begin{split}
A_{\text{w}}B_{\text{w}}C_{\text{w}}A_{\text{w}}B_{\text{w}}C_{\text{w}}A_{\text{c}}B_{\text{c}}C_{\text{c}}A_{\text{c}}B_{\text{c}}C_{\text{c}}...\\
......A_{\text{c}}B_{\text{c}}C_{\text{c}}C_{\text{w}}B_{\text{w}}A_{\text{w}}C_{\text{w}}B_{\text{w}}A_{\text{w}}
\end{split}
\end{equation}
will lead to long range deformation of the crystal.

In Fig.~\ref{fig:11} and its inset, we plot  $\gamma_{\text
{wc}}$ as a function of the structured wall density and, in the inset,
as a function of the wall-crystal interaction strength $\epsilon_{\rm pw}$. 
Similar to the liquid case we find
that the interfacial free energy decreases with $\epsilon_{\rm pw}$ due
to the stronger attraction between the crystal and the wall. Unlike the
liquid case, Fig.~\ref{fig:11} shows that while the main
trend for $\gamma_{\rm wc}$ is to increase with decreasing density
of the structured wall, there is a sharp dip when the density
of the wall equals the density of the crystal. This is easy
to understand, since less energy will be needed to create an interface,
when the structured wall has the same structure as the crystal than when
there is a mismatch between the wall and crystal structures leading to
a relatively unfavorable interaction between them.

\section{Conclusion}
\label{sec_conc}
We propose a thermodynamic integration (TI) scheme to compute interfacial
free energies of liquids or crystals in contact with flat or structured
walls from molecular dynamics simulation. In this work, this scheme has
been applied to Lennard-Jones systems, but it can be easily generalized
to other interaction models. The implementation of our method is simple,
and, as demonstrated above, our method provides reliable and accurate
estimates of $\gamma_{\rm wl}$ and $\gamma_{\rm wc}$ that enter in
Young's equation (\ref{eq:young}).  In particular for structured walls
(substrates), to the best of our knowledge, there are no simulation
studies calculating the substrate-crystal interfacial free energy
$\gamma_{\rm wc}$.  Most of the previous simulation works on structured
walls \cite{nijmeijer90,tang95,vanleeuwen90,bruin91,bakker89,crevecoeur95}
have been limited to the calculation of the interfacial free energy
$\gamma_{\rm wl}$ using the integration over the pressure anisotropy (PA).
The PA method, however, does not give reliable results in general, and,
in contrast to our TI scheme, it is not applicable to substrates that
can support stress (such as structured walls where the wall particles are
allowed to move and are thus not fixed to their ideal lattice positions,
see discussion above).  Therefore, the TI scheme proposed in this work
can be considered as a novel approach to obtain accurate values for
substrate-liquid or substrate-crystal interfacial free energies and thus
it will be useful in studies of wetting and nucleation problems.

\begin{acknowledgments}
One of the authors (R. B.) thanks the DLR-DAAD fellowship program for
financial support.  The authors acknowledge financial support by the
German DFG SPP 1296.  Computer time at the NIC J\"ulich is gratefully
acknowledged.
\end{acknowledgments}

\end{document}